\documentclass[12pt]{article}
\usepackage[centertags]{amsmath}
\usepackage{amsfonts}
\usepackage{amssymb}
\usepackage{amsthm} 
\usepackage{newlfont}
\usepackage{epsfig}
\usepackage{amscd}
\usepackage{graphicx}
\usepackage{epsfig}
\usepackage{amscd}
\usepackage{showlabels}
\usepackage{color}

\newcommand{\beq}{\begin{equation}}
\newcommand{\eeq}{\end{equation}}
\newcommand{\ba}{\begin{array}}
\newcommand{\ea}{\end{array}}
\newcommand{\bea}{\begin{eqnarray}}
\newcommand{\eea}{\end{eqnarray}}
\begin{document}

\begin{center}
{\large \sc \bf {Informational correlation between two parties of a quantum system:
spin-1/2 chains. }
}

\vskip 15pt

{\large 
 A.I.~Zenchuk 
}

\vskip 8pt

{\it Institute of Problems of Chemical Physics, RAS,
Chernogolovka, Moscow reg., 142432, Russia, e-mail:   zenchuk@itp.ac.ru
 } 
\end{center}

%\today

\begin{abstract}
We introduce the  informational  correlation $E^{AB}$ 
between two interacting quantum subsystems $A$ and $B$ of a quantum system  as the number of arbitrary parameters $\varphi_i$ of a   unitary  transformation $U^A$ (locally performed on  the subsystem $A$)  which may be detected in the subsystem $B$ 
by the local measurements.  This quantity indicates whether  the state of the subsystem $B$ may be effected by means of the unitary transformation applied to  the subsystem $A$. Emphasize that $E^{AB}\neq E^{BA}$ in general. 
The informational correlations in systems with tensor product initial states are studied in more details. 
In particular,  it is shown that  the informational correlation may be changed by the local unitary transformations of the subsystem $B$. However, there is some non-reducible part  of  $E^{AB}(t)$ which may not be decreased by any unitary  transformation of the subsystem $B$ at a fixed  time instant $t$.  
Two examples of the informational correlations between two parties of
the  four node spin-1/2  chain with mixed initial states are studied. The long chains with a
single initially excited spin (the pure initial state) are considered as well.

\end{abstract}

%%%%%%%%%%%%%%%%
\section{Introduction}
\label{Sec:Introduction}

Quantum correlations are considered  to be responsible for advantages of  quantum
devices in comparison with their classical analogues. To quantify these advantages, a special measures of 
quantum correlations have been introduced. 
The  entanglement  \cite{W,HW,P,AFOV,HHHH} and  discord \cite{HV,OZ,Zurek,L,ARA}
are known as two basic measures. However, the role of such correlations is not completely clarified. Of course, they have to  be available  in the system. But,  up to now, it is not clear whether the quantum correlations  must be large in order to reveal all advantages of quantum  information devises. 
Moreover, there are many verifications of the hypothesis that the quantum correlations  measured either 
by entanglement or discord shouldn't be large.  For instance, there are quantum states without entanglement revealing  a quantum non-locality \cite{BDFMRSSW,HHHOSSS,NC}. In addition,   
speed-up of certain calculations may be organized with negligible entanglement \cite{M,DFC,DV,DSC,LBAW}. 

{{}
We also should remember that the discord is not the only measure of quantum correlations showing
that entanglement does 
not cover all of them. For instance, the so-called localizable entanglement (LE) \cite{VPC,JK}  
was introduced as a maximal amount of entanglement which can be established  between two selected 
particles of a quantum system by doing   local measurements on the
rest of the quantum system. In ref.\cite{JK}, considering examples of the 
anti-ferromagnetic spin chains authors show that the local measurements result 
in increase of the bi-particle 
entanglement. In some sense, LE  is the "intermediate" measure of quantum correlations 
between the entanglement  and discord. In the later case, the local measurements are performed on  one of 
two selected  particles. The advantage of LE is that its upper and lower bounds  are directly related to 
 correlation functions which, in turn,  have a physical nature.} 

These interesting results together with an observation that almost all quantum states
possess the non-vanishing quantum discord 
\cite{FACCA} might lead us to the conclusion (which perhaps causes doubts)  
that almost all quantum systems may be effectively used in the  quantum information devices.   
 
 Thus, the above {{} discussion} suggests us to conclude that even the quantum devices based on 
 the states  with  minor (but nonzero) entanglement and/or discord   may 
 exhibit advantages in comparison with their classical counterparts. Such behavior   may be  explained by the  observation that the spread of  information about the state of a given subsystem throughout the  whole  quantum system does not require the large values of either   entanglement or discord \cite{Z_Inf}.
 In other words, if we change a state of a given subsystem $A$ at the initial time instant $t=t_0$ (for instance, applying the unitary transformation),
 then, generally speaking,  any other subsystem   of the  whole quantum system will know about the new state  of $A$ at (almost)  any instant $t>t_0$.  In turn, namely this information provides the overall mutual relations  among all parties of a quantum system. Moreover, 
it is valuable that the spread (or distribution) of information throughout the whole system   does not require the fine adjustment of the parameters of a quantum system, unlike the large discord and/or entanglement, when the minor deviation of the system's parameters 
from the required values may crucially  decrease the values of these measures. It is worthwhile to say that the spread of information is observed even in the system with separable initial state when there is neither discord no entanglement between subsystems initially \cite{Z_Inf}. Of course the discord and entanglement  may  appear in the course of evolution, but their values are not crucial for the information distribution. The measure of correlations introduced  
in this paper is based on the outlined above information distribution. 
 
Before proceed to the subject of our paper, let us notice that the system-environment states  with  
vanishing  discord give impact to 
  study  the evolution of a system  as a completely positive map \cite{BP} from the initial state 
  of this  system to its evolving state   \cite{RMKSS,SL,BDKMRR}.  However, 
  although originally the completely positive maps were found for the states with   initially
 vanishing discord \cite{RMKSS,SL}, it was shown later that  the vanishing discord is not necessary  
 for this \cite{BDKMRR}. Thus, here the situation is opposite to the quantum nonlocality and speedup.
 
 Similar to the above refs.\cite{RMKSS,SL,BDKMRR}, we consider the evolution of the state of a given  
 subsystem $B$  of some quantum system as a map of  the initial state of another subsystem $A$ and show that 
  this  map is responsible  for 
 the information distribution  between two  chosen subsystems. Then,  correlations (which absent initially) appear as result of such evolution.
 Our basic study is referred to  the systems 
 having the tensor product initial states (and zero discords), when the above map becomes completely positive one. However, our algorithm may be extended to  systems with  general initial states (this problem is briefly discussed in Appendix C, Sec.\ref{Section:C}).

 In this paper, we do not separate quantum and classical effects.
  Instead, we  study the 
 possibility to handle the state of some subsystem $B$ at some instant $t$ 
by means of the local unitary transformation $U^A$ of another  subsystem $A$   at instant $t_0<t$. 
We refer to  the measure quantifying this effect  as  {\it the informational correlation} $E^{AB}$
between two parties $A$ and $B$ of a quantum system.  
In this regard, one has to mention Ref.\cite{MCD} where the quantumness of  operations 
has been studied without the direct relation to the entanglement and discord, similar to our case. 

 We also shall  note, that although the discord and entanglement
 reveal the physical nature of quantum correlations, they may not be 
  directly used as "working units" in  quantum devices, because, in general,
  they are not expressed in terms of measurable   
  quantities. On the contrary, the informational correlation is expressed  in terms of the parameters 
  of unitary transformation showing how many of these parameters may be 
  transfered between two subsystems. These parameters, in turn, 
  might be treated, for instance, as bits of quantum gates.
  This conclusion is a basic motivation for the
  introduction of the informational correlation.
  
Unlike the usual definition of the information through the entropy \cite{NCbook},
we define the measure of  informational correlation  
as the number of  parameters $\varphi_i$ of the local unitary transformation $U^A(\varphi^A)$, $\varphi^A=\{\varphi_1,\varphi_2,\dots\}$ that may be detected in the subsystem $B$ by means of the local measurements.   Vise-verse,
the  influence of the subsystem $B$  on the subsystem $A$ may be characterized by the informational correlation  
$E^{BA}$ which equals  the number of  parameters $\varphi_i$ in the local transformation $U^B(\varphi^B)$, $\varphi^B=
\{\varphi_1,\varphi_2,\dots\}$, of the subsystem $B$   that may be detected in the subsystem $A$ by means of the local   measurements. 
We assume that namely these measures $E^{AB}$ and  $E^{BA}$  characterize the  strength of those quantum correlations which are responsible for the  information exchange between parties. In turn, namely the mutual exchange of the parameters  may be used in the realization of elementary logical gates.  
Emphasize that, in general, ${\mbox{dim}} A\neq {\mbox{dim}} B$ and $E^{AB}\neq E^{BA}$, i.e., this measure is not symmetrical, similar to the discord, which depends on the particular subsystem chosen for the local classical  measurements.  
For the tensor product initial state $\rho(0)=\rho^A(0)\otimes \rho^{C}(0)\otimes \rho^B(0)$
(where $\rho^A$ and $\rho^B$ are  respectively the density matrices of the subsystems  $A$ and $B$,  while
$\rho^{C}$ is the density matrix of the rest of the quantum system), it will be shown  that $E^{AB}= E^{BA}=0$  only if both initial density matrices of subsystems $A$ and $B$ are proportional to the unit matrix. Note, that the unitary invariant discord \cite{Z_Disc} possesses the same property. 

Below, we  study the informational correlation $E^{AB}$  only. 
For the tensor product initial state $\rho(0)=\rho^A(0)\otimes \rho^{CB}(0)$ (where $\rho^{CB}$ is the density matrix of the subsystem $C\cup B$), it will be shown, in particular,  that the informational correlation is invariant with respect to the initial local  unitary transformations of the subsystem $A$ (i.e., $E^{AB}$ 
depends only on the eigenvalues of the initial density matrix $\rho^A(0)$)   and might be changed  by the initial local  unitary transformations of  the subsystem $B$. 
In addition, we reveal such part of the informational correlation which may not be decreased  by the local unitary transformations of  the subsystem $B$ at a given time instant $t$ (the so-called non-reducible informational correlation). 

{{}
Among the local unitary transformations one should select a special type of local operations (the so-called 
locally non-effective unitary operations (LNUs)) which do 
not change the state of the subsystem $A$ but may change the 
state of the whole quantum system. 
The effect of these transformations  on the state of the whole system 
has been studied in refs.\cite{F,GKB,DG}. However, using  LNUs, 
we may obtain only zero informational correlation 
in a system with the tensor product initial state considered in our paper.
In other words, LNUs  are useless in our case.
However, we assume the relevance of such transformations for systems with
more general initial states. Such systems  deserve the special study and are
not  considered here  in details
(some notes about these systems  are given in  Appendix C, Sec.\ref{Section:C}).

In this paper the relation of the informational correlation with the well known measures of quantum correlations 
 (such as entanglement and discord) is  discussed only on the  rather qualitative background and 
 there is no direct analogy between them.  The most principal difference is the fact that the informational 
 correlation defined in our paper is a dynamical quantity and   equals to zero identically 
 at the initial time instant
 (i.e., at the instant of applying of the local unitary transformation to the subsystem $A$).
We consider the tensor product  initial state, 
 so that  both entanglement and discord equal to zero initially. 
Thus, we represent the detailed analysis of informational correlation in the systems with 
initially zero  entanglement/discord, 
while the systems with other initial states (where the  entanglement and/or discord are, 
generally speaking, non-zero) deserve the special study. }

The paper is organized as follows.
In Sec.\ref{Section:def}, we introduce  the definition of the informational 
correlation and discuss the non-reducible informational correlation. Examples 
of the informational correlations in the four node spin-1/2 homogeneous chain 
governed by the XY Hamiltonian with the nearest neighbor interactions are 
considered in Sec.\ref{Section:examples}.
Particular examples of the informational correlations in a long spin chain with single excitation 
are considered in Sec.\ref{Section:long}.  In Sec.\ref{Section:conclusions},
we collect the basic properties of the informational correlation and 
compare them with the corresponding properties of the entanglement/discord. 
Some additional information and calculations are given
in the Appendix, Sec.\ref{Section:appendix}. {{}
In particular, Appendix C (Sec.\ref{Section:C}) is devoted to some properties 
of informational correlation in the systems with non-separable initial states.}

\section{Definition of informational correlation}
\label{Section:def}

As is noted in Sec.\ref{Sec:Introduction}, the definition of informational correlation is based on 
the parameters of unitary transformation. To proceed with,  we introduce the 
set of  notations for a given quantum system $S$.  
\begin{eqnarray}\label{unitary}
&&
U^S \;\;\;{\mbox{is the unitary transformation performed on the system $S$}},\\
\label{dim}
&&
N^S={\mbox{dim}}\,S \;\;{\mbox{is the dimensionality of the system}},\\
\label{MS}
&&  M^S=N^S(N^S-1)/2 \;\;{\mbox{is the number of off-diagonal elements in}} \\\nonumber
 && \hspace{2cm} {\mbox{the $N^S\times N^S$  matrix}},\\
\label{DAt}
&&
D^S=(N^S)^2-1 \;\;{\mbox{is the total number of arbitrary  parameters}}\\\nonumber
&& \hspace{4cm} {\mbox{parameterizing the group $SU(N^S)$}},\\
\label{varphiS}
&&
\varphi^S=\{\varphi_1,\dots,\varphi_{D^S}\}\;\;{\mbox{is the set of parameters parameterizing the group $SU(N^S)$}},\\\nonumber
&&\hspace{1cm} 
\varphi^S \in   \bar G^S
,
\end{eqnarray}
where $ \bar G^S$ is the closed region in the space of the parameters $\varphi_i$, $i=1,\dots, D^S$. As usual, 
$  G^S$ denotes the appropriate open region. 
Hereafter we assume that the whole quantum system is splitted into the three subsystems $A$, $B$ and $C$. We study the correlations 
between the subsystems  $A$ and $B$, while $C$ is the rest of a quantum system. Thus, the total system is $A\cup C \cup B$. In particular, the subsystem $C$ may be absent.
Let the state of the whole quantum system be described by the density matrix $\rho$. In turn, 
as usual, the states of the subsystems $A$, $B$, $C$ and $C\cup B$  are represented by the reduced 
density matrices $\rho^A$, $\rho^B$, $\rho^C$ and $\rho^{CB}$ respectively:
\begin{eqnarray}
\rho^A={\mbox{Tr}}_{BC} \rho,\;\;\;\rho^B={\mbox{Tr}}_{AC} \rho,\;\;\;\rho^C={\mbox{Tr}}_{AB} \rho,\;\;\;\rho^{CB}={\mbox{Tr}}_{A} \rho.
\end{eqnarray}

{{}The measure of correlations proposed below is based on the effects of the local 
unitary transformations performed on  one of two selected  subsystems of a quantum system.
Let us}  effect on the state of the subsystem $B$ by means of the unitary transformation $U^A(\varphi^A)$ of the subsystem $A$ at the initial instant $t = 0 $.   Let us determine how many parameters of the arbitrary transformation $U^A(\varphi^A)\in SU(N^A)$ may be detected in the subsystem $B$ at this instant. 
For this purpose we, first of all, fix the 
  state of the system at $t=0$  by the initial density matrix $\rho(0)$.
The local transformation $U^A(\varphi^A)\in SU(N^A)$   transforms the initial  density matrix  $\rho(0)$ of the whole system  $A\cup C\cup B$ into the density matrix  $\rho(\varphi^A,0)$ 
as follows:
\begin{eqnarray}\label{rho_phi_A}
 \rho(\varphi^A,0)=(U^A(\varphi^A)\otimes I_C \otimes I_B) \rho(0) ((U^A(\varphi^A))^+\otimes I_C\otimes I_B).
\end{eqnarray}
Now the initial density matrix $\rho^A(\varphi^A,0)$ at $t=0$  reads
\begin{eqnarray}\label{UA0}
\rho^A(\varphi^A,0) ={\mbox{Tr}}_{BC}  \rho(\varphi^A,0) =  U^A(\varphi^A) \rho^A(0) (U^A(\varphi^A))^+,
\end{eqnarray}
while the initial density matrix $\rho^B(\varphi^A,0)$ remains the same,
\begin{eqnarray}\label{UB0}
\rho^B(\varphi^A,0) ={\mbox{Tr}}_{AC} \rho(\varphi^A,0) = \rho^B(0) ,
\end{eqnarray}
which means that no parameters $\varphi_i$ may be detected in $\rho^B$ at $t=0$.

Eqs.(\ref{UA0}) and (\ref{UB0}) are valid 
for any subsystems $A$ and $B$ no matter whether there is quantum interaction  between them.
Next, if the subsystems $A$ and $B$ do not interact, then no information about the state of the subsystem 
$A$ propagates into the subsystem $B$  \cite{Z_Inf}. Consequently, no parameters $\varphi_i$ of 
the applied transformation $U^A(\varphi^A)$ may be detected in the subsystem $B$. 
In other words,   the performed transformation will not effect on the state of the  subsystem $B$.

However, the information about the state of the subsystem $A$ propagates into the subsystem $B$  if there is quantum interaction between these subsystems. Owing to this interaction, some of the parameters of the unitary transformation $U^A(\varphi^A)$ may be  transfered  into the subsystem $B$. This interaction is represented by the unitary $t$-dependent transformation applied to  the whole system and leads to the evolution of the density matrix:
\begin{eqnarray}\label{ev}
\rho(\varphi^A,t)=V(t) \rho(\varphi^A,0) V^+(t) .
\end{eqnarray}
In particular, if  the evolution of our quantum system  is  governed by the stationary  Hamiltonian ${\cal{H}}$, then 
\begin{eqnarray}\label{Wt}
V(t)=\exp\Big(
- i {\cal{H}} t
\Big)
\end{eqnarray}
in accordance with the Liouville equation.
In this case, the state of the subsystem $B$ is represented by the following  reduced evolution  density matrix:
\begin{eqnarray}\label{Ut}
\rho^B(\varphi^A,t) = {\mbox{Tr}}_{AC} \left(V(t) \Big(U^A(\varphi^A)\otimes I_C\otimes I_B\Big) \rho(0) 
\Big((U^A(\varphi^A))^+\otimes I_C \otimes I_B \Big) V^+(t)\right) \neq \rho^B(0).
\end{eqnarray}
Hereafter we consider the initial density matrix $\rho(0)$ in the form of the tensor product of two initial density matrices:
\begin{eqnarray}\label{inden}
\rho(0) =\rho^A(0) \otimes \rho^{CB}(0)  ,
\end{eqnarray}
where $\rho^A(0) $  and $\rho^{CB}(0)$  are  the initial density matrices  
of the subsystem $A$  and of  the joined subsystem  $C\cup B$, respectively.
Now, {{} using eqs.(\ref{rho_phi_A}) and (\ref{inden}), we may write eq.(\ref{Ut}) 
 as
\begin{eqnarray}\label{Ut2}\label{TTmel}
\rho^B(\varphi^A,t) = {\mbox{Tr}}_{AC} \left(V(t) \Big(\rho^A(\varphi^A,0)\times \rho^{CB}(0)\Big) 
V^+(t)\right).
\end{eqnarray}
}
\iffalse
in terms of the matrix elements as follows:
\begin{eqnarray}\label{TTmel}
\rho^B_{i_B; j_B}(\varphi^A,t)&=&
\sum_{i_A , k_A, n_A=1}^{N^A} 
\sum_{i_C , k_C, n_C=1}^{N^C} 
\sum_{ k_B, n_B=1}^{N^B}
V_{i_Ai_Ci_B;k_Ak_Ck_B}(t)  \rho^A_{k_A ;n_A}(\varphi^A,0) 
\times\\\nonumber
&&
\rho^{CB}_{k_Ck_B; n_Cn_B}(0)  V^+_{n_An_Cn_B;i_Ai_Cj_B}(t).
\end{eqnarray}
\fi
For the subsequent analysis, 
we separate the $t$- and $\varphi^A$-dependences  in the rhs of  eq.(\ref{TTmel}) 
into two different matrices.{{} This structure would 
be more convenient for the study of the $\varphi_i$-dependence 
of both lhs and rhs of eq.(\ref{TTmel}). Doing this, we first write eq.(\ref{TTmel}) 
in components 
using the following matrix representations of  the operators: 
$\rho^A(0) =\{\rho^A_{k_A;n_A}(0)\}$,  $\rho^{CB}(0)=\{\rho^{CB}_{k_Ck_B;n_Ck_C}(0)\}$,
$V=\{ V_{i_Ai_C i_B;n_A k_Ck_B}\}$);
the indices with subscripts $A$, $B$ and $C$ are related to the subsystems $A$, $B$ and $C$  respectively.
 Writing the matrix representation we 
may use any  basis associated with a proper subsystem.
For instant, in the case of spin-1/2 chains considered below this basis might be composed by the eigenstates of the 
$z$-projection operators of  spin angular momenta. Thus, we have the following:} 
\begin{eqnarray}\label{TTmel2}
\rho^B_{i_Bj_B}(\varphi^A,t) = \sum_{k_A,n_A=1}^{N^A}  T_{i_Bj_B;k_An_A}(t)\rho^A_{k_A;n_A}(\varphi^A,0),\;\;
i_B,j_B=1,\dots,N^B,
\end{eqnarray}
where the  elements $T_{i_Bj_B;n_Am_A}$ {{}(independent  on $\varphi^A$)} are defined by the formulas:
\begin{eqnarray}\label{T}
T_{i_Bj_B;n_Am_A}(t) &=& \sum_{i_A =1}^{N^A} 
\sum_{i_C , k_C, n_C=1}^{N^C} 
\sum_{ k_B, n_B=1}^{N^B}
V_{i_Ai_C i_B;n_A k_Ck_B}(t) 
\rho^{CB}_{k_C k_B; n_B n_C}(0)  V^+_{m_A n_Cn_B;i_Ai_Cj_B}(t),\\\nonumber
&&
i_B,j_B=1,\dots,N^B,\;\;n_A,m_A=1,\dots,N^A.
\end{eqnarray}
The important feature of eq.(\ref{TTmel2}) is that the $\varphi^A$-dependence appears only through the initial 
density matrix $\rho^A(\varphi^A,0)$ of the subsystem $A$. This becomes possible because of
the tensor product initial state (\ref{inden}). 
Such separation of $t$- and $\varphi^A$-dependences is impossible  for other initial states,
see Appendix C, Sec.\ref{Section:C}.

Further,  in order to determine the number of parameters transfered from $A$ to $B$, 
it is convenient to represent  eq.(\ref{TTmel2}) in a different form.
The matter is that eqs.(\ref{TTmel2}), as well as the density matrices $\rho^A$ and $\rho^B$, 
are complex while the parameters $\varphi_i$ are real. Therefore we {{}
represent  eqs.(\ref{TTmel2}) as a  system of real equations. For this purpose, we split eq.(\ref{TTmel2})
into the real and imaginary parts  writing the result in terms of the real 
and imaginary parts of the density matrices $\rho^A$ and $\rho^B$ \cite{Z_Inf}.
Doing this, we need  the following  notations (hereafter we use the matrix indices without the subscripts $A$, $B$, $C$)}:
\begin{eqnarray}\label{hTT1}
&&
T^1_{ij;nm}=T_{ij;nm}+T_{ij;mn},\;\;T^2_{ij;nm}=T_{ij;nm}-T_{ij;mn},\\\nonumber
&&
i,j = 1,\dots, N^B,\;\;n,m=1,\dots,N^A,\;\;m> n.
\end{eqnarray}
Now, we write eq.(\ref{TTmel2}) as  the following three subsystems: 
\begin{eqnarray}\label{RIT1}
{\mbox{Re}} \rho^B_{ij}(\varphi^A,t)& =& \sum_{{m,n=1}\atop{m>n}}^{N^A} \left(
{\mbox{Re}} T^1_{ij;nm} {\mbox{Re}} \rho^A_{nm}(\varphi^A,0) -
{\mbox{Im}} T^2_{ij;nm} {\mbox{Im}} \rho^A_{nm}(\varphi^A,0) 
\right) +\\\nonumber
&&
\sum_{n=1}^{N^A} {\mbox{Re}} (T_{ij;nn}) \rho^A_{nn}(\varphi^A,0),\;\;i,j=1,\dots,N^B, \;\; j>i\\\label{RIT2}
{\mbox{Im}} \rho^B_{ij}(\varphi^A,t) &=& \sum_{{n,m=1}\atop{m>n}}^{N^A} \left(
{\mbox{Im}} T^1_{ij;nm} {\mbox{Re}} \rho^A_{nm}(\varphi^A,0) +
{\mbox{Re}} T^2_{ij;nm} {\mbox{Im}} \rho^A_{nm}(\varphi^A,0) 
\right) +\\\nonumber
&&
\sum_{n=1}^{N^A} {\mbox{Im}} (T_{ij;nn}) \rho^A_{nn}(\varphi^A,0),
\;\;i,j=1,\dots,N^B,\;\;j>i,
\\\label{RIT3}
 \rho^B_{ii}(\varphi^A,t) &=& \sum_{{n,m=1}\atop{m>n}}^{N^A} \left(
{\mbox{Re}} T^1_{ii;nm} {\mbox{Re}} \rho^A_{nm}(\varphi^A,0) -
{\mbox{Im}} T^2_{ii;nm} {\mbox{Im}} \rho^A_{nm}(\varphi^A,0) 
\right) +\\\nonumber
&&
\sum_{n=1}^{N^A} T_{ii;nn} \rho^A_{nn}(\varphi^A,0),\;\;i=1,\dots,N^B-1,
\end{eqnarray}
where subsystem (\ref{RIT1}) is the real off-diagonal part of eq.(\ref{TTmel2}), subsystem 
(\ref{RIT2}) is the imaginary off-diagonal part of the same equation,  and subsystem (\ref{RIT3})
is the diagonal (real) part of eq.(\ref{TTmel2}). We also take into account the 
relation ${\mbox{Tr}} \rho^B =1$, which leaves $N^B-1$ {{} (instead of $N^B$)} 
independent equations in subsystem  (\ref{RIT3}).
Therewith elements $T^k_{ij;nm}$, $k=1,2$, possess the following symmetry with respect to 
the indices $i$ and $j$:
\begin{eqnarray}\label{hTT2}
T^1_{ij;nm} = (T^1_{ji;nm})^*, \;\;T^2_{ij;nm} = -(T^2_{ji;nm})^*, \;\;T_{ij;nn} = T^*_{ji;nn},
\end{eqnarray}
where star means the complex conjugate.
{{} However, the structure of system (\ref{RIT1}-\ref{RIT3}) is still 
not the best one for the further computations. To provide an elegant expression for 
the informational correlation, we  represent system 
(\ref{RIT1}-\ref{RIT3}) as a single vector equation. For this purpose, we relate  three vectors $X(\rho)$, 
$Y(\rho) $ and
$Z(\rho) $ 
with 
 any $N\times N$ density matrix $\rho$. The components of these vectors 
 $X_\alpha(\rho)$, $Y_\alpha(\rho)$, $\alpha=1,\dots, M$, $M\equiv N(N-1)/2 $, 
and  $Z_i(\rho)$, $i=1,\dots,N-1$ are defined as}
%\sum_{l=1}^{i-1} (N-l)+j-i
\begin{eqnarray}\label{XYZ}
&&
X_{\sum_{l=1}^{i-1} (N-l)+j-i}(\rho)= {\mbox{Re}}\, \rho_{ij},\;\;i,j=1,\dots,N,\;\;j>i,\\\nonumber
&&
Y_{\sum_{l=1}^{i-1} (N-l)+j-i}(\rho)= {\mbox{Im}}\, \rho_{ij},\;\;i,j=1,\dots,N,\;\;j>i,\\\nonumber
&&
Z_{i}(\rho)=  \rho_{ii},\;\;i=1,\dots,N-1.
\end{eqnarray}
{{} Using these vectors, we rewrite 
eqs.(\ref{RIT1}-\ref{RIT3}) in the following forms:}
\begin{eqnarray}\label{RITex1}
X_\alpha(\rho^B(\varphi^A,t)) &=& \sum_{\beta=1}^{M^A} \Big(
 T^{11}_{\alpha\beta}(t) X_\beta(\rho^A(\varphi^A,0)) +
 T^{12}_{\alpha\beta}(t) Y_\beta(\rho^A(\varphi^A,0)) 
\Big) +\\\nonumber
&&
\sum_{n=1}^{N^A-1} (T^{13}_{\alpha n}(t) -T^{13}_{\alpha N^A}(t)) Z_n(\rho^A(\varphi^A,0))+ T^{13}_{\alpha N^A}(t),\;\;\alpha =1,\dots M^B,\\\nonumber
Y_\alpha(\rho^B(\varphi^A,t)) &=& \sum_{\beta=1}^{M^A} \Big(
 T^{21}_{\alpha\beta}(t) X_\beta(\rho^A(\varphi^A,0)) +
 T^{22}_{\alpha\beta}(t) Y_\beta(\rho^A(\varphi^A,0)) 
\Big)+\\\nonumber
&&
\sum_{n=1}^{N^A-1} (T^{23}_{\alpha n}(t) -T^{23}_{\alpha N^A}(t)) Z_n(\rho^A(\varphi^A,0))+ T^{23}_{\alpha N^A}(t),\;\;\alpha =1,\dots M^B,
\\\nonumber
 Z_i(\rho^B(\varphi^A,t)) &=& \sum_{\beta=1}^{M^A} \Big(
 T^{31}_{i\beta}(t) X_\beta(\rho^A(\varphi^A,0)) +
 T^{32}_{i\beta}(t) Y_\beta(\rho^A(\varphi^A,0)) 
\Big) +\\\nonumber
&&
\sum_{n=1}^{N^A-1} (T^{33}_{in}(t)-T^{33}_{iN^A}(t)) Z_n(\rho^A(\varphi^A,0)) + T^{33}_{iN^A}(t),\;\;i=1,\dots,N^B-1,
\end{eqnarray}
where we introduce the  matrices $T^{ij}$ with the  elements:
\begin{eqnarray}\label{hTT3}
&&
T^{11}_{\sum_{l=1}^{i-1} (N^B-l)+j-i, \sum_{l=1}^{n-1} (N^A-l)+m-n} = {\mbox{Re}} T^1_{ij;nm},\\\nonumber
&&
T^{12}_{\sum_{l=1}^{i-1} (N^B-l)+j-i, \sum_{l=1}^{n-1} (N^A-l)+m-n} = -{\mbox{Im}} T^2_{ij;nm},\;\;
%\\\nonumber
%&&
T^{13}_{ \sum_{l=1}^{i-1} (N^B-l)+j-i ,n} = {\mbox{Re}} T_{ij,nn},
\\\nonumber
&&
T^{21}_{\sum_{l=1}^{i-1} (N^B-l)+j-i, \sum_{l=1}^{n-1} (N^A-l)+m-n} = {\mbox{Im}} T^1_{ij;nm},\\\nonumber
&&
T^{22}_{\sum_{l=1}^{i-1} (N^B-l)+j-i,\sum_{l=1}^{n-1} (N^A-l)+m-n} = {\mbox{Re}} T^2_{ij;nm},\;\;
T^{23}_{ \sum_{l=1}^{i-1} (N^B-l)+j-i ,n} = {\mbox{Im}} T_{ij,nn},\\\nonumber
&&
T^{31}_{i, \sum_{l=1}^{n-1} (N^A-l)+m-n} = {\mbox{Re}} T^1_{ii;nm},\;\;
T^{32}_{i,  \sum_{l=1}^{n-1} (N^A-l)+m-n} =- {\mbox{Im}} T^2_{ii;nm},\;\;
T^{33}_{ i ,n} = T_{ii,nn}.
\end{eqnarray}
Thus $T^{11}$,   $T^{12}$,  $T^{21}$,  $T^{22}$ are $M^B\times M^A$ matrices, 
$T^{13}$ and $T^{23}$ are $M^B\times N^A$ matrices, $T^{31}$ and $T^{32}$ are  $(N^B-1)\times M^A$
matrices and $T^{33}$ is $(N^B-1)\times N^A$ matrix. 
Next, we construct the column vectors  $\hat X(\rho^A)$ and  $\hat X(\rho^B)$ of $D^{A} = (N^A)^2 -1$ 
and $D^{B} = (N^B)^2 -1$ elements respectively, and  the column vector $\hat T^0(t)$ of $D^{B}$ elements:
\begin{eqnarray}\label{XT}
&&
\hat X(\rho^A(\varphi^A,0)) =\left(
\begin{array}{c}
X(\rho^A(\varphi^A,0))\cr
Y(\rho^A(\varphi^A,0))\cr
Z(\rho^A(\varphi^A,0))
\end{array}
\right),
\;\;\hat X(\rho^B(\varphi^A,t)) =\left(
\begin{array}{c}
X(\rho^B(\varphi^A,t))\cr
Y(\rho^B(\varphi^A,t))\cr
Z(\rho^B(\varphi^A,t))
\end{array}
\right),\\\label{XT2}
&&
\hat T^0(t) = \left(
\begin{array}{c}
T^{10}(t)\cr
T^{20}(t)\cr
T^{30}(t)
\end{array}\right),
\end{eqnarray}
Here the elements of  the vectors $T^{k0}(t)$, $k=1,2,3$, are  defined as 
$T^{i0}_\alpha(t)= T^{i3}_{\alpha,N}$, $i=1,2$, $\alpha=1,\dots,M^B$, and 
  $T^{30}_i(t)= T^{33}_{i,N}$, $i=1,\dots,N^B-1$.
We also introduce  the $D^B\times D^A$ block matrix $\hat T(t)$:
\begin{eqnarray}\label{hTT4}
\hat T(t) = \left(
\begin{array}{ccc}
T^{11}(t)  & T^{12}(t) &\tilde T^{13}(t) \cr
T^{21}(t)  & T^{22}(t) &\tilde T^{23}(t)\cr
T^{31}(t)  & T^{32}(t) &\tilde T^{33}(t) 
\end{array}
\right),
\end{eqnarray}
where  the elements of the matrices 
$\tilde T^{k3}(t)$ are defined as $\tilde T^{k3}_{\alpha n} =
 T^{k3}_{\alpha n} -T^{k3}_{\alpha N^A}$, $k=1,2$,  $\tilde T^{33}_{in} =
 T^{33}_{i n} -T^{33}_{i N^A}$, $\alpha=1,\dots,M^B$, 
 $i=1,\dots,N^B-1$, $n=1,\dots, N^A-1$.
Finally, system (\ref{RITex1}) {{} becomes the} single 
$((N^A)^2-1)\times 1$ vector equation:
\begin{eqnarray}\label{RITf}
\hat
X(\rho^B(\varphi^A,t)) = 
\hat T(t)
\hat
X(\rho^A(\varphi^A,0)) +\hat T^0(t).
\end{eqnarray}
 {{} 
 Thus, we derive eq.(\ref{RITf}) as the most convenient form of  original 
 operator equation (\ref{TTmel}). Namely eq.(\ref{RITf}) will be used hereafter for the 
 calculation of  informational correlation and for study of its properties. }
 
 {{}
 \paragraph{Complete information transfer and maximal
 value of informational correlation $E^{AB}$.}
 Eq.(\ref{RITf}) shows that 
 the vector $X(\rho^A(\varphi^A,0))$ may be uniquely expressed in terms of the 
 elements of the vector $X(\rho^B(\varphi^A,t))$ (describing the 
 measured state of the subsystem $B$) if $\hat T$ is an invertible square matrix \cite{Z_Inf}. 
 In other words, all elements of  the density matrix $\rho^A(\varphi^A,0)$ 
can be transfered into the subsystem $B$ in this case (the complete information transfer). 
However, we show in this paragraph that the number of  parameters encoded into the matrix 
$\rho^A(\varphi^A,0)$
is less than the length of the column $\hat X$. Thus it is natural to assume that 
the complete information transfer is not required in order to transfer all 
encoded parameters $\varphi_i$  from $A$ to $B$, i.e., the maximal value for $E^{AB}$ 
is achievable without the complete information transfer.}

 {{}Let us calculate the number $\tilde D^A$ of   arbitrary parameters encoded 
into the matrix $\rho^A(\varphi^A,0)$. }
 For this purpose  we  represent an  arbitrary matrix $\rho^A(0)$ in the form
\begin{eqnarray}
\rho^A(0)= W \Lambda^A W^+, \;\;W\in SU(N^A),
\end{eqnarray}
where $\Lambda^A={\mbox{diag}}\{\lambda^A_1,\dots,\lambda^A_{N^A}\}$ is the diagonal matrix of the eigenvalues and $W$ is the matrix of eigenvectors of   $\rho^A(0)$. Then we may write  
\begin{eqnarray}\label{rhoA}
\rho^A(\varphi^A,0) = U^A(\varphi^A)W \Lambda^A W^+ (U^A(\varphi^A))^+= \tilde U^A(\tilde \varphi^A) 
\Lambda^A (\tilde U^A(\tilde\varphi^A))^+\equiv \rho^A(\tilde\varphi^A,0) ,
\end{eqnarray}
where $\tilde U^A(\tilde \varphi^A) = U^A(\varphi^A) W \in SU(N^A )$, and 
$\tilde\varphi^A=\tilde \varphi^A(\varphi^A) =\{\tilde \varphi_1,\dots,\tilde \varphi_{D^A}\}$ is the set of   redefined (in terms of $\varphi_i$) parameters of the group $SU(N^A)$.
Now, we  calculate the number of arbitrary  parameters $\tilde \varphi_i$ encoded 
into the matrix $\rho^A(\tilde \varphi^A,0)$ as follows. The maximal possible  number of the independent real parameters in the $N^A\times N^A$ dimensional matrix $\rho^A$ is $(N^A)^2-1$. But $N^A-1$ of them are related with the  eigenvalues of the density matrix (the $N^A-1$ diagonal elements of 
the matrix $\Lambda^A$ in eq.(\ref{rhoA}) {{} taking} into account the relation ${\mbox{Tr}}\rho^A(0) =
\sum_{i=1}^{N^A} \lambda^A_i \equiv 1$). These elements are fixed  {{} by 
the initial density matrix $\rho^A(0)$.} Consequently, we stay with 
\begin{eqnarray}\label{DA}
\tilde D^A = (N^A)^2-1 -(N^A-1)=N^A(N^A-1)
\end{eqnarray}
arbitrary real parameters in the density matrix $\rho^A(\varphi^A,0)$. These parameters 
are related with the same number of  parameters $\tilde \varphi_i$ in the 
transformation $\tilde U(\tilde\varphi^A)$. Thus, only $\tilde D^A$ arbitrary parameters 
of the group $SU(N^A)$ may be encoded into the density matrix $\rho^A(\tilde\varphi^A)$. 
{{}The value of $\tilde D^A$ (\ref{DA})  
is less then the length of the vector $\hat X(\rho^A(\tilde \varphi^A,0))$ which is 
$((N^A)^2-1)$. 
This means that not all elements of the density matrix $\rho^A(\tilde\varphi^A,0)$ must be 
transfered into the matrix $\rho^B(\tilde\varphi^A,t)$ in order to detect all $\tilde D^A$ 
parameters $\tilde\varphi_i$ in the subsystem $B$. Therefore, the complete 
information transfer \cite{Z_Inf}, in principle, is not required in order 
to transfer the maximal possible number of arbitrary parameters $\tilde\varphi_i$ from 
the subsystem $A$ into the subsystem $B$.}
Hereafter we  consider only the diagonal matrix $\rho^A(0)$ without loss of generality  
and do not write the tilde over $\varphi^A$.

\paragraph{
{{}The number of parameters encoded into the subsystem $A$.}
}
{{} We see that } the  parameter $\tilde D^A$ 
(determined  by eq.(\ref{DA})) indicates the maximal 
possible number of parameters $\varphi_i$ encoded into the matrix $\rho^A(\varphi^A,0)$. 
However, this maximum is not always achievable, which suggests us to introduce another quantity  
$E^{AA}$ indicating the number of 
parameters $\varphi_i$ which are really encoded into the state of the subsystem $A$. This number is completely defined 
by the multiplicity of the eigenvalues $\lambda_i$. In fact,
if all $\lambda_i$ are different, then the number of parameters encoded 
into the subsystem $A$ is $\tilde D^A$, 
i.e., $E^{AA}=\tilde D^A$. 
Now we assume that  there is  one $K$-fold eigenvalue, $K\le N^A$. Then $\Lambda^A$ is invariant with respect to the proper group  $SU(K)$ which possesses $d= K ( K-1)$ parameters. These $d$ parameters may not be encoded into $\rho^A$. Consequently,  the number of  encoded  parameters $E^{AA}$ 
becomes less then $\tilde D^A$:
\begin{eqnarray}\label{tDK0}
E^{AA}=\tilde D^A-K(K-1).
\end{eqnarray}
Formula (\ref{tDK0}) may be readily extended to the case of  $Q$ roots with multiplicities $K_i>1$, $i=1,\dots,Q$:
\begin{eqnarray}\label{tDK}
E^{AA}=\tilde D^A-\sum_{i=1}^Q K_i(K_i-1),\;\; \sum_{i=1}^Q K_i \le N^A.
\end{eqnarray}
This formula for $E^{AA}$ will be used in Sec.\ref{Section:examples}.

\paragraph{Zero informational correlations, $E^{AB}=E^{BA}=0$, in system with tensor product 
initial state (\ref{inden}).} 
Thus, we  measure the informational correlation  by the number $E^{AB}$ of  arbitrary  parameters 
$\varphi_i$ of the unitary transformation $U^A(\varphi^A) \subset SU(N^A)$  which may be deduced 
from the analysis of the matrix $\rho^B(\varphi^A,t)$ describing the state of the   subsystem $B$.
If all $\lambda_i$ are the same 
(i.e., $\Lambda^A$ is proportional to the unit matrix) then  $\rho^A(\varphi^A,0)$ 
is also proportional 
to the unit matrix. Consequently  no parameters $\varphi_i$ 
may be encoded into $\rho^A(\varphi^A,0)$. 
Therefore,  no parameters of the unitary transformation  $U^A$ may be transfered to the subsystem $B$, 
i.e., $E^{AB} =0$. However,   the parameters might be transfered in the opposite direction (from $B$ to $A$).
In fact,  let us assume that 
\begin{eqnarray}\label{inden0}
\rho^{CB}(0) = \rho^C(0)\otimes \rho^B(0)
\end{eqnarray}
for simplicity. If not all  eigenvalues of the initial density matrix $\rho^B(0)$ are the same (i.e., the matrix $\rho^B(0)$ is not proportional 
to the identity matrix),
then at least  some of  parameters of the unitary transformation of the subsystem $B$  
may be encoded into the density matrix $\rho^B(\varphi^B,0)$.  {{} 
Then, these parameters may be } transfered to the subsystem $A$ (although this still depends on the $t$-evolution operator), i.e., $E^{BA}\neq 0$.  
Thus, no parameters may be transfered in both directions if only both $\rho^A(0)$ and 
$\rho^B(0)$ are proportional to the identity matrix. Emphasize that this conclusion holds only for the  tensor product  initial state $\rho(0)=\rho^A(0) \otimes \rho^C(0) \otimes\rho^B(0)$. 
Note that  the unitary invariant discord \cite{Z_Disc} is zero in such systems as well.  
The  zero informational correlations  in systems with more general initial states are not considered in this paper.

%%%%%%%%%%%%%%%%
\subsection{Calculation of the parameters $E^{AA}$ and $E^{AB}$}
\label{Section:EAAB}
As mentioned above {{}(see the paragraph after eq.(\ref{DA})),} 
we  take the diagonal initial density matrix $\rho^A(0)$,
\begin{eqnarray}
\rho^A(0) \equiv \Lambda^A,\;\;\;
\lambda^A_1\ge \lambda^A_2\ge \dots\ge \lambda^A_{N^A},
\end{eqnarray}
without loss of generality.
Obviously, $E^{AB}$ may not exceed  $E^{AA}$ (the number of parameters $\varphi_i$ encoded into the density 
matrix $\rho^A(\varphi^A,0)$; {{}this parameter is introduced above, see  eqs.(\ref{DA}-\ref{tDK})
and the text therein)} 
$E^{AA}\le \tilde D^A$. In turn, $E^{AA}$ is 
uniquely defined by the multiplicity of the eigenvalues of the density matrix $\rho^A(0)$ (see eq.(\ref{tDK})). 
Let us calculate the informational correlation $E^{AB}$ following its definition as the number 
of arbitrary parameters transfered from the subsystem $A$ to the subsystem $B$. This number equals to the number of parameters $\varphi_i$ which might be found from  vector eq.(\ref{RITf}) with known  lhs (the matrix $\rho^B$ in the lhs must be determined by the local measurements). 
This equation is a {{}vector} transcendental equation, whose global solution  may not be given analytically. However, 
we may   define the number of  different detectable parameters  in the close 
neighborhood of any fixed point $\varphi^A\in G^A$. This is  the number of  functionally independent 
elements  of  the vector $\hat X(\rho^B(\varphi^A,t))$, which, in turn,  equals to the rank of the  Jacobian  matrix,
\begin{eqnarray}\label{JA}
J(\rho^B(\varphi^A,t))=\frac{\partial (\hat X_1(\rho^B(\varphi^A,t)),\dots, 
\hat X_{(N^B)^2-1}(\rho^B(\varphi^A,t))  }{\partial(\varphi_1,\dots,\varphi_{\tilde D^A})}, 
\end{eqnarray}
calculated in the above fixed point $\varphi^A \in  G^A$.
Therefore,  we determine the informational correlation as 
\begin{eqnarray}\label{EAB}
E^{AB}(\varphi^A,t) = {\mbox{ran}}\, J(\rho^B(\varphi^A,t)).
\end{eqnarray}
Similarly, {{}
the  parameter  $E^{AA}$ } may be determined as  the  rank of the Jacobian matrix $J(\rho^A(\varphi^A,t))$,
\begin{eqnarray}\label{JB}
J(\rho^A(\varphi^A,t))=\frac{\partial (\hat X_1(\rho^A(\varphi^A,t)),\dots, 
\hat X_{(N^A)^2-1}(\rho^A(\varphi^A,t))  }{\partial(\varphi_1,\dots,\varphi_{\tilde D^A})},
\end{eqnarray}
as follows:
\begin{eqnarray}\label{EAA}
E^{AA}(\varphi^A) =
{\mbox{ran}} \; J(\rho^A(\varphi^A,0)).
\end{eqnarray}
Moreover, we may readily write the relations between two Jacobian matrices  
$J(\rho^B(\varphi^A,t))$ and $J(\rho^A(\varphi^A,0))$ differentiating eq.(\ref{RITf}) with respect to the parameters $\varphi_i$:
\begin{eqnarray}\label{Jf}
J(\rho^B(\varphi^A,t)) = 
\hat T(t)
J(\rho^A(\varphi^A,0)).
\end{eqnarray}
From eq.(\ref{Jf}) in virtue of eq.(\ref{EAA})  it follows that
\begin{eqnarray}\label{Jfneq}
E^{AB}(\varphi^A,t) \le \min \Big({\mbox{ran}}\, \hat T(t),
E^{AA}(\varphi^A) \Big),\;\;\varphi^A\in G^A.
\end{eqnarray}
All in all, 
it  is demonstrated  that the informational correlation
 $E^{AB}(\varphi^A,t)$ depends on  two factors. 
\begin{enumerate}
\item
The  number of  arbitrary parameters $\varphi_i$  which might be   encoded into the density matrix $\rho^A(\varphi^A,0)$ (quantity 
 $E^{AA}$).
\item
The  number of  arbitrary parameters which can be  transfered from the subsystem $A$ to the subsystem $B$. If the information is completely transfered, then $E^{AB} = E^{AA}$.
Otherwise $E^{AB} \le E^{AA}$.
\end{enumerate}

Note that  $E^{AA}$ defined by eq.(\ref{EAA}) does not really depend on $\varphi^A$ (if only  
$\varphi^A \in G^A$) owing to the mutually unique map $\rho^A(\varphi^A,0) \leftrightarrow \varphi^A$. 
This unique map means that, for a given set of  eigenvalues of $\rho^A(0)$, 
we have  the appropriate number of the independent elements of the vector 
$\varphi^A$, which uniquely parametrize the matrix $\rho^A(\varphi^A,0)$ (\ref{UA0}).
This number equals to the number of functionally independent elements in 
the vector $X(\rho^A(\varphi^A,0))$. The later  equals to  the rank of the Jacobian matrix $J(\rho^A(\varphi^A,0))$ and must be the same for all $\varphi^A$ 
at least inside of $G^A$, where $J(\rho^A(\varphi^A,0))$ is well defined. 

Regarding the informational correlation $E^{AB}(\varphi^A,t)$ given by  eq.(\ref{EAB}),  it really depends on 
$\varphi^A$ in the case of general initial state $\rho(0)$, as  shown in Appendix C, Sec.\ref{Section:C}. However, regarding initial state (\ref{inden}), we may readily conclude  that $E^{AB}$  does not depend on $\varphi^A$, $\varphi^A \in G^A$. 
The reason is  that initial state  (\ref{inden}) results in the separation of $t$- and $\varphi^A$-dependence in  eq.(\ref{Jf}) relating $J(\rho^B(\varphi^A,t))$ with 
$J(\rho^A(\varphi^A,0))$. It is important that $\varphi^A$-dependence is concentrated in $J(\rho^A(\varphi^A,0))$. 
Therefore,  multiplying $J(\rho^A(\varphi^A,0))$  by $\hat T(t)$ we only recombine rows of 
the matrix $J(\rho^A(\varphi^A,0))$. Consequently, the rank of the resulting matrix $J(\rho^B(\varphi^A,t))$ 
does not depend on $\varphi^A$, but it depends on $t$. For this reason,  we will not write $\varphi^A$ in the arguments of  $E^{AB}$ and $E^{AA}$ (except for the Appendix C, Sec.\ref{Section:C}), i.e.,
\begin{eqnarray}\label{novarphi}
E^{AA}(\varphi^A) \equiv E^{AA},\;\;\; 
E^{AB}(\varphi^A,t)\equiv E^{AB}(t).
\end{eqnarray}
 {{} 
Consequently, $E^{AB}$ is defined by the  eigenvalues of the density matrix  $\rho^A(0)$  (which
do not depend on $\varphi_i$)
rather then by its elements themselves, 
which is similar to the unitary invariant discord  \cite{Z_Disc}. }

Two simple examples of informational correlations in the 4-node spin-1/2 chain will be considered in Sec.\ref{Section:examples}.

%%%%%%%%%%%%%%%%%%%%
\paragraph{Effect of the local initial  unitary transformation of the subsystems 
$C$ and $ B$ on  $E^{AB}$.}

While the initial local transformations of the subsystem 
$A$ do not  effect the informational correlation (they only 
lead to  the redefinition of the independent parameters $\varphi_i$),
the initial local unitary transformation of the subsystems $C$ and $B$
may change  the informational correlation $E^{AB}$. 
In this paper we study only    the diagonal initial state of the subsystems 
$C$ and $B$
%, 
 %$\rho^{CB}(0)={\mbox{diag}}(\lambda^{CB}_1,\dots,\lambda^{CB}_{N^{CB}})$, 
 and demonstrate the effect of the local  unitary transformations
 {{}of these subsystems} considering 
 the particular examples in Sec.\ref{Section:loc}  and in the end of 
 Sec.\ref{Section:ranT2}.

%%%%%%%%%%%%%%%%%%%%
\paragraph{Normalization of  informational correlation.}

As  mentioned  above {{}(see eq.(\ref{Jfneq}))}, the informational correlation $E^{AB}(t)$ in a system
with tensor product initial state (\ref{inden}) may not exceed  
the maximal possible number $\tilde D^A$ (\ref{DA}) of the parameters 
$\varphi_i$ encoded into the density matrix $\rho^A(\varphi^A,0)$. To indicate the discrepancy between  
$E^{AB}(t)$ and $\tilde D^A$, we introduce the normalized measure $E^{AB}_{norm}$ as the ratio
\begin{eqnarray} \label{norm}
E^{AB}_{norm}= \frac{E^{AB}}{\tilde D^A}.
\end{eqnarray}
Thus, the maximal value of $E^{AB}_{norm}$ is unit at least in the   quantum systems with  the tensor product 
initial state (\ref{inden}), for which inequality (\ref{Jfneq}) holds. This is, generally speaking, not valid  in the  case of an arbitrary initial state, see Appendix C, Sec.\ref{Section:C}.
Similarly, we normalize $E^{AA}$:
\begin{eqnarray} \label{normA}
E^{AA}_{norm}= \frac{E^{AA}}{\tilde D^A}.
\end{eqnarray}

%%%%%%%%%%%%%%%%%
\subsection{Non-reducible informational correlation}
\label{Section:nonred}
It is noted in the end of Sec.\ref{Section:EAAB} that the informational correlation 
is sensitive to the initial unitary transformation of the subsystem $B$.
Moreover, it is simple to show  that
the informational correlation $E^{AB}(t)$ determined at some instant $t$ may 
be decreased by the local unitary transformation of the subsystem $B$ at the same   instant $t$,
unlike the entanglement and discord.  In fact, 
let us represent the density matrix $\rho^B(\varphi,t)$  in the form  
\begin{eqnarray}\label{BUL}
&&
\rho^B(\varphi^A,t) = U^B(\varphi^A,t) \Lambda^B(\varphi^A,t) (U^B(\varphi^A,t))^+,\\\label{BULlam}
&&
\Lambda^B={\mbox{diag}}(\lambda_1,\dots,\lambda_{N^B}),
\end{eqnarray}
where $\Lambda^B$ is the diagonal matrix of the eigenvalues and $U^B$ is composed of the eigenvectors of the matrix $\rho^B(\varphi^A,t)$. 
Note that we do not write the superscript $B$ in the notation $\lambda_i$ to defer these eigenvalues from 
the eigenvalues of the initial density matrix $\rho^B(0)$ considered in Sec.\ref{Section:examples}.  
Representation (\ref{BUL}) suggests us to  split the whole set of  transfered parameters into two subsets. 
The first subset collects  those parameters which may be detected from the analysis of the matrix $U^B$  at instant $t$ (the subset $\varphi^U$), while the second one collects those parameters which  may be detected from the analysis of the matrix $\Lambda^B$ at the same time instant (the subset $\varphi^\Lambda$). 
Therewith some of the parameters might appear in both subsets. Other parameters might not appear in these subsets at all. So, 
$E^{AB}$ equals the number of different parameters in two subsets $\varphi^U$ and $\varphi^\Lambda$. 
Moreover, eq.(\ref{BUL}) shows that one can eliminate parameters $\varphi^U$ from the reduced  density matrix
$\rho^B(\varphi^A,t)$ at instant $t$  performing the local unitary transformation $ (U^B(\varphi^A,t))^+$, which transforms the matrix $\rho^B$ to the diagonal form,
\begin{eqnarray}\label{map}
\rho^B(\varphi^A,t) \to \Lambda^B(\varphi^A,t).
\end{eqnarray}
 {{} This reduces} the set of  transfered parameters to $\varphi^\Lambda$. {{} Consequently} the informational correlation $E^{AB}$ 
reduces  to $E^{AB;min}(\varphi^A,t)$ which  equals to 
the number of parameters in $\varphi^\Lambda$.  Of course, by definition, this part of  
the informational correlation   may not be decreased by any local unitary transformation of  
the subsystem $B$ at instant $t$. We refer to $E^{AB;min}(\varphi^A,t)$ as {\it the non-reducible informational correlation}.
Obviously, the following upper estimation is valid:
\begin{eqnarray}\label{upper}
E^{AB;min}(\varphi^\Lambda,t) \le N^B-1,
\end{eqnarray}
because there are only $N^B-1$ independent eigenvalues owing to the relation ${\mbox{Tr}}\rho^B =1$. Similarly to $E^{AB}$ and $E^{AA}$ (see eqs.(\ref{EAB}) and (\ref{EAA}) respectively), the non-reducible informational correlation $E^{AB;min}(\varphi^A,t) $ may be calculated as 
 the rank of the Jacobian matrix $J^B_\Lambda(\varphi^A,t)$,
\begin{eqnarray}
J^B_\Lambda(\varphi^A,t)=\frac{\partial (\lambda_1(\varphi^A,t),\dots, \lambda_{N^B-1}(\varphi^A,t))}
{\partial(\varphi_1,\dots,\varphi_{\tilde D^A})},
\end{eqnarray}
i.e., by the formula
\begin{eqnarray}\label{nonredE}
E^{AB;min}(\varphi^A,t) = {\mbox{ran}}\,J^B_\Lambda(\varphi^A,t),
\end{eqnarray}
where
 $\lambda_i(\varphi^A,t)$, $i=1,\dots,N^B-1$ are the independent eigenvalues. 
This correlation may be also normalized:
\begin{eqnarray}
E^{AB;min}_{norm}(\varphi^A,t) = \frac{E^{AB;min}(\varphi^A,t)}{\tilde D^A}.
\end{eqnarray}

Let us  {{} derive} the more applicable form of eq.(\ref{nonredE}) in terms of the principal minors of the matrix $\rho^B(\varphi^A,t)$. 
We start with the characteristic equation for the matrix $\rho^B(\varphi^A,t)$:
\begin{eqnarray}
\label{char0}
&&
\prod_{i=1}^{N^B} (\lambda-\lambda_i) =0,\;\;\;{\mbox{or}}\\
\label{char}
&&
\lambda^{N^B} -\lambda^{N^B-1} +\sum_{i=0}^{N^B-2} a_i(\varphi^A,t) \lambda^i =0,\;\;
\end{eqnarray}
where
\begin{eqnarray}
a_0(\varphi^A,t)=(-1)^{N^B} \det\, \rho^B(\varphi^A,t), \;\;a_i(\varphi^A,t)=(-1)^{N^B-i} S_{N^B-i}(\varphi^A,t),\;\;i=1,\dots ,N^B-2,
\end{eqnarray}
and $S_j$ means the sum of all the $i$th-order principal minors of the matrix $\rho^B(\varphi^A,t)$.  In eq.(\ref{char}),  we take into account that ${\mbox{Tr}}\rho^B=1$. 
Differentiating eq.(\ref{char})  with respect to the parameter $\varphi_k$ and solving the resulting equation for $\displaystyle \frac{\partial\lambda}{\partial \varphi_k}$ we obtain:
\begin{eqnarray}
\frac{\partial\lambda}{\partial \varphi_k} =
\frac{-\displaystyle \sum_{i=0}^{N^B-2} \frac{\partial a_i(\varphi^A,t) }{\partial\varphi_k} \lambda^i}{\displaystyle N^B\lambda^{N^B-1} - (N^B-1) \lambda^{N^B-2}+ \sum_{i=1}^{N^B-2}  i a_i(\varphi^A,t) \lambda^{i-1}}.
\end{eqnarray}
Therefore, the matrix $J^B_\Lambda(\varphi^A,t)$ reads
\begin{eqnarray}\nonumber
&&\hspace{-1cm}
J^B_\Lambda(\varphi^A,t)=
{J_0^{-1}(\varphi^A,t)}\left(
\begin{array}{ccc}\displaystyle
\sum_{i=0}^{N^B-2}  \frac{\partial a_i(\varphi^A,t)}{\partial\varphi_1} \lambda^i_1(\varphi^A,t)&\cdots&\displaystyle
\sum_{i=0}^{N^B-2}  \frac{\partial a_i(\varphi^A,t)}{\partial\varphi_{\tilde D^A}} \lambda^i_{1}(\varphi^A,t)\cr
\cdots&\cdots&\cdots \cr\displaystyle
\sum_{i=0}^{N^B-2}  \frac{\partial a_i(\varphi^A,t)}{\partial\varphi_1} \lambda^i_{N^B_1-1}(\varphi^A,t)&\cdots&\displaystyle
\sum_{i=0}^{N^B-2}  \frac{\partial a_i(\varphi^A,t)}{\partial\varphi_{\tilde D^A}} \lambda^i_{N^B-1}(\varphi^A,t)
\end{array}
\right) =\\\label{LamH}
&&J_0^{-1}(\varphi^A,t)\hat \Lambda^B(\varphi^A,t)H(\varphi^A,t) ,
\end{eqnarray}
where 
\begin{eqnarray}
J_0(\varphi^A,t)&=&(-1)^{N^B-1}
\prod_{j=1}^{N^B-1} \Big(N^B\lambda^{N^B-1}_j(\varphi^A,t) - (N^B-1) \lambda^{N^B-2}_j(\varphi^A,t)+ \\\nonumber
&&
\sum_{i=1}^{N^B-2}  i a_i(\varphi^A,t) \lambda^{i-1}_j(\varphi^A,t)\Big) \neq 0,
\end{eqnarray}
while  $\hat\Lambda^B$ and $H$  are the $(N^B-1)\times (N^B-1)$ and  $ (N^B-1)\times \tilde D^A$ matrices respectively:
\begin{eqnarray}\label{hLambda}
&&
\hat\Lambda^B(\varphi^A,t)=\left(
\begin{array}{cccc}\displaystyle
  1&\lambda_1 (\varphi^A,t)&\cdots&\displaystyle
  \lambda_1^{N^B-2} (\varphi^A,t)\cr
  \cdots&\cdots&\cdots&\cdots\cr
  1&\lambda_{N^B-1} (\varphi^A,t)&\cdots&\displaystyle
  \lambda_{N^B-1}^{N^B-2} (\varphi^A,t)
\end{array}
\right),
\\
\label{HH}
&&
H(\varphi^A,t)=\left(
\begin{array}{ccc}\displaystyle
  \frac{\partial a_0(\varphi^A,t)}{\partial\varphi_1} &\cdots&\displaystyle
  \frac{\partial a_0(\varphi^A,t)}{\partial\varphi_{\tilde D^A}} \cr
\cdots&\cdots&\cdots \cr\displaystyle
 \frac{\partial a_{N^B-2}(\varphi^A,t)}{\partial\varphi_1} &\cdots&\displaystyle
  \frac{\partial a_{N^B-2}(\varphi^A,t)}{\partial\varphi_{\tilde D^A}} 
\end{array}
\right).
\end{eqnarray}
If all $\lambda_i$ ($i=1,\dots,N^B-1$)  are different and nonzero, then $\det \hat \Lambda^B\neq 0$ and 
\begin{eqnarray}\label{JH}
E^{AB;min}(\varphi^A,t) ={\mbox{ran}} \,J^B_\Lambda(\varphi^A,t)={\mbox{ran}} \,H(\varphi^A,t).
\end{eqnarray}
 It is not difficult to prove  that eq.(\ref{JH}) holds even for the case of  multiple and/or zero  eigenvalues $\lambda_i$, $i=1,\dots,N^B-1$, which is shown in the Appendix B, Sec.\ref{Section:B}. 
 
 Emphasize that, unlike $E^{AB}$ and $E^{AA}$,  the non-reducible 
correlation $E^{AB;min}(\varphi^A,t)$ depends on $\varphi^A$ even for the product initial state (\ref{inden}).
This means that there might be such points  $\varphi^A_1$  and   $\varphi^A_2$  in the space of $\varphi^A$ that  $E^{AB;min}(\varphi^A_1,t) \neq E^{AB;min}(\varphi^A_2,t)$.
However, there might exist such $g\subset G^A$  and appropriate  time intervals $T_{1i}<t<T_{2i}$, $i=1,2,\dots$ 
that   $E^{AB;min}$ is independent on $\varphi^A\in g$ inside of {{}these} 
time intervals. 
In this case, it might be reasonable  to write $E^{AB;min}(g,t)$ instead of 
$E^{AB;min}(\varphi^A,t)$, see Secs.\ref{Section:nonredEX1} and \ref{Section:nonredEX2}. 
{{} In addition, it is quite possible that $g = G^A$. 
Namely this situation is realized in
Sec.\ref{Section:nonredEX1}.}

The non-reducible correlation may be also normalized:
\begin{eqnarray}
E^{AB;min}_{norm}(\varphi^A,t) = \frac{E^{AB;min}(\varphi^A,t)}{\tilde D^A}.
\end{eqnarray}

A  particular example  of single parameter in the subset $\varphi^\Lambda$ is considered analytically in Sec.\ref{Section:nonredEX1}. 
In general, the calculation of $E^{AB;min}$ is a numerically solvable problem, 
which is partially  discussed in the Appendix C, Sec.\ref{Section:C}, and in   two examples of
Sec.\ref{Section:nonredEX2}.

 %%%%%%%%%%%%%%%%
 \subsection{Removable informational correlation}
 
 The non-reducible informational correlation is the analogue of classical correlations in the calculation of  discord \cite{HV,OZ,Zurek}.
 However, we do not identify 
 the non-reducible informational correlation with the classical part of the  informational correlation, because $E^{AB;min}$ might be related with some  quantum effects as well. Now, having $E^{AB}$ and $E^{AB;min}$, we define {\it the  removable informational correlation} as the increment 
 \begin{eqnarray}\label{incr}
 \Delta E^{AB}(\varphi^A,t) = E^{AB}(t)-E^{AB;min}(\varphi^A,t).
 \end{eqnarray}
 This correlation may be normalized as
 \begin{eqnarray}\label{incr_norm}
 \Delta E^{AB}_{norm}(\varphi^A,t) =\frac{\Delta E^{AB}(\varphi^A,t)}{\tilde D^A} = E^{AB}_{norm}(t)-E^{AB;min}_{norm}(\varphi^A,t).
 \end{eqnarray}
Since the  removable informational correlation  is defined by  those parameters $\varphi_i$ which may be detected only from the matrix $U^B(\varphi^A,t)$ at instant $t$,
it may be considered as an analogue of the quantum correlations in the calculation of discord
\cite{HV,OZ,Zurek} (in other words, $\Delta E^{AB}$ is  the analogue of discord). However,  we do not state that this measure characterizes the pure quantum effects. 
 
 {{}
 \subsection{Informational correlation established by the subgroup of $SU(N^A)$}
 In this subsection we briefly outline  the case when the 
 informational correlation is established by a 
 subgroup of $SU(N^A)$. Using a subgroup we reduce the number of parameters $\varphi_i$
 involved in the process. As a consequence, both the number of encoded parameters 
 $E^{AA}$ and the informational correlation $E^{AB}$ become reduced. However, 
 the system under such 
 transformations may reveal some peculiarities. In particular, the study of informational 
 correlations induced by the LNUs \cite{F,GKB,DG} would be of interest. However, LNUs 
 are useless for a system with the  tensor product  initial state (\ref{inden}), 
 because LNUs do not effect 
 the state of the whole system in this case. For this reason we do
 not consider LNUs in this  paper. These transformations must be studied for the systems with
 the non-separable initial states (see Appendix C, Sec.\ref{Section:C}), 
 that will be done in a different paper.
 }
 
%%%%%%%%%%%%%%%%%%%%
\section{Four-node homogeneous spin-1/2 chain}
\label{Section:examples}
We consider the spin-1/2 system of four nodes 
whose evolution is governed by the  $XY$ Hamiltonian
\begin{eqnarray}\label{H}
{\cal{H}}= -\sum_{i=1}^{3} \frac{d}{2} (I^+_iI^-_{i+1} + I^-_iI^+_{i+1})
\end{eqnarray}
with the nearest neighbor interaction. 
Here  $d$ is the coupling constant between the nearest neighbors, 
$I^\pm_i=I_{x;i} \pm i I_{y;i}$ and $I_{\alpha;i}$, $\alpha=x,y,z$, are the projection operators  of the total spin angular momentum. 
We put $d=1$ without the loss of generality. Hamiltonian (\ref{H}) must be used in the  evolution operator  $V(t)\in SU(16)$ defined by eq. (\ref{Wt}).

%%%%%%%%%%%%%%%
\subsection{One-node subsystems $A$ and $B$}
\label{Section:N4}
Let the subsystems $A$  and $B$ be  represented by the first and the last nodes, respectively, while  
the subsystem $C$  consists of two middle  nodes. Thus $N^A=N^B=2$ and $N^C=4$, so that $SU(2)$ is the 
group of unitary  transformations of the subsystem $A$.

We consider the  initial density matrix  having the structure 
(\ref{inden}) with the tensor product matrix $\rho^{CB}(0)$ given by eq.(\ref{inden0}).
{{} So,} the initial density matrix $\rho(0)$ reads
\begin{eqnarray}\label{rho0ex1}
\rho(0)= \rho^{A}(0) \otimes\rho^{C}(0) \otimes \rho^{B}(0).
\end{eqnarray}
{{} According to } Sec.\ref{Section:EAAB}, 
we take the diagonal initial density matrix $\rho^A(0)$ 
{{}without the loss of generality,}
\begin{eqnarray}
\rho^A(0)={\mbox{diag}}(\lambda^A,1-\lambda^A),
\end{eqnarray}
and choose the diagonal matrices $\rho^B(0)$ and $\rho^C(0)$ as well:
\begin{eqnarray}\label{N4rho01}
\rho^B(0)={\mbox{diag}}(\lambda^B,1-\lambda^B),\;\;\;
 \rho^C(0)={\mbox{diag}}(\lambda^C_1,\lambda^C_2,\lambda^C_3,1-\lambda^C_1-\lambda^C_2-\lambda^C_3).
\end{eqnarray}

\subsubsection{Number of parameters encoded into the subsystem $A$, $E^{AA}$}
\label{Section:EAA1n}
The general form of $SU(2)$ transformation 
reads \cite{G_book}
\begin{eqnarray}\label{UA}
U^A(\varphi^A) &=&\left( \begin{array}{cc}
\cos\varphi_1 &-e^{-i \varphi_2}\sin\varphi_1 \cr
e^{i \varphi_2}\sin\varphi_1 &\cos\varphi_1
\end{array}
\right)
e^{i \sigma_3 \varphi_3},\;\;\sigma_3={\mbox{diag}}(1,-1),
\\\label{int}
&&G^A: \;\;0 < \varphi_1,\varphi_3 < \frac{\pi}{2},\;\;\;0 < \varphi_2 <  2\pi,
\end{eqnarray}
where we do not consider the boundary values of the parameters $\varphi_i$, $i=1,2,3$.
Since $\rho^A(0)$ is diagonal, the parameter $\varphi_3$ does not appear 
in $\rho^A(\varphi^A,0)$. Therefore only two parameters can be encoded into 
the density matrix $\rho^A(\varphi^A,0)$, i.e., $\tilde D^A=2$. The same result 
may be obtained directly using the formula (\ref{EAA}) with the Jacobian matrix 
given  by formula (\ref{JB}). {{}Therewith} 
the vector $X(\rho^A(\varphi^A,0))$, defined by eqs.(\ref{XYZ},\ref{XT}), reads
\begin{eqnarray}
\hat X(\rho^A(\varphi^A,0))&=&  
\{{\mbox{Re}} \rho^A_{12}(\varphi,0), {\mbox{Im}} \rho^A_{12}(\varphi^A,0), 
\rho^A_{11}(\varphi^A,0)\}^T=\\\nonumber
&&
\{
(2\lambda^A-1) \sin\,\varphi_1 \cos\,\varphi_1 \cos\,\varphi_2,\\\nonumber
&&-(2\lambda^A-1) \sin\,\varphi_1 \cos\,\varphi_1 \sin\,\varphi_2,
\\\nonumber
&&
\frac{1}{2}(1+(2\lambda^A-1) \cos(2 \varphi_1))  
\}^T.
\end{eqnarray}
Here  the superscript $T$ means the matrix transposition. 
Thus, the parameter $\varphi_1$ appears in three entries of the column $\hat X$, while $\varphi_2$ 
appears only in two  entries. This observation suggests us to consider the parameter $\varphi_1$ as a 
more reliable one  {{} for  realization  of} the informational correlation. In fact, it will be shown in Sec.\ref{Section:nonredEX1} that namely this parameter 
is responsible for the non-reducible informational correlation between the subsystems $A$ and $B$ (i.e., between the first 
and the  last nodes of the 4-node spin chain).
All in all we {{} obtain the values 
for $E^{AA}$ (and $E^{AA}_{norm}$) collected in Table 1.}
\iffalse
\begin{eqnarray}\label{A1_0}
E^{AA}=\left\{\begin{array}{ll}
2,&\displaystyle \lambda^A\neq \frac{1}{2},\cr
0,&\displaystyle \lambda^A=\frac{1}{2}
\end{array}\right..
\end{eqnarray}
\fi

\subsubsection{Relation between  the rank of  $\hat T$ and the informational correlation $E^{AB}$}
\label{Section:ranT}
Now we turn to the whole quantum system $A\cup C \cup B$ and consider the evolution of the density matrix
 $\rho(\varphi^A,t)$ of this system,
\begin{eqnarray}\label{rhot}
&&
\rho(\varphi^A,t) = V(t) \rho(\varphi^A,0)  V^+(t) ,
\end{eqnarray}
where
\begin{eqnarray}
\rho(\varphi^A,0)= \Big(\tilde U^A(\varphi^A) \times I_4\times I_2\Big) \rho(0)
\Big( ( \tilde U^A(\varphi^A))^+ \times I_4\times I_2\Big) .
\end{eqnarray}
The evolution operator $ V(t)\in SU(16)$ is given by eq.(\ref{Wt}), and   $\rho(0)$ is given by eq.(\ref{rho0ex1}).
To calculate $E^{AB}$ we refer to eqs.(\ref{JA} -- \ref{Jf}) and {{} write}  all matrices used in these equations.
The  vector $X(\rho^B(\varphi^A,t))$   associated with 
the local density matrix $\rho^B(\varphi^A,t)={\mbox{Tr}}_{AC} \; \rho(\varphi^A,t) $
  is defined by eqs.(\ref{XYZ},\ref{XT}) as follows:
\begin{eqnarray}\label{Xrho}
X(\rho^B(\varphi^A,t)) &=&
\{{\mbox{Re}} \rho^B_{12}(\varphi^A,t), {\mbox{Im}} \rho^B_{12}(\varphi^A,t),
\rho^B_{11}(\varphi^A,t)\}^T
\end{eqnarray}
{{}(we do not write the explicit formulas for the rhs of eq.(\ref{Xrho}) because they are very cumbersome).}
The matrices $\hat T$ and $\hat T^0$ read:
\begin{eqnarray}\label{hT}
\hat T(t)=\left(\begin{array}{ccc}
0 & a_{1}(t) &0\cr
-a_{1}(t) & 0&0\cr
0&0&a_2(t),
\end{array}\right),\;\;
\hat T^0(t)=\left(\begin{array}{c}
0\cr
0\cr
b(t),
\end{array}\right),
\end{eqnarray}
where
\begin{eqnarray}\label{a1a2b}
a_{1}(t) &=& 
\frac{(2 \lambda^B-1)(2 \lambda^C_3+
2 \lambda^C_2 -1)}{5+\sqrt{5}} r(t),\\\nonumber
\\\nonumber
a_2(t)&=& \frac{r(t)^2}{10(3+\sqrt{5}) }
\\\nonumber
b(t)&=&\lambda^B\left(
\frac{3+2\cos\frac{\sqrt{5} t}{2} }{5 }- a_2(t)\right)+
\frac{2\sin^2\frac{\sqrt{5}t}{4}}{5} 
\Big(
2\lambda^C_1 +\lambda^C_2+ \lambda^C_3  +  (\lambda^C_3-\lambda^C_2)
\cos\frac{t}{2}
\Big)
\end{eqnarray}
with
\begin{eqnarray}\label{r}
r(t)=2\sin\frac{(1+\sqrt{5})t}{4} + (3 +\sqrt{5})
\sin\frac{(1-\sqrt{5})t}{4} .
\end{eqnarray}
Formulas (\ref{a1a2b}) for the entries of the matrix $\hat T$ suggest us to consider only such time instants that satisfy the condition 
\begin{eqnarray}\label{rneq0}
r(t)\neq 0,
\end{eqnarray}
because otherwise the rank of $\hat T$ is zero. 
The first positive root of the expression in the lhs of condition (\ref{rneq0})  is $t=9.070$. Thus, a suitable time interval for the parameter detection in the subsystem $B$ might be the following one: $0<t<9.070$.
Under  condition (\ref{rneq0}), 
eqs.(\ref{hT}) and (\ref{a1a2b}) yield
\begin{eqnarray}\label{ran}
{\mbox{ran}} \,\hat T(t) =\left\{
\begin{array}{ll}
3 ,&\displaystyle \lambda^B \neq \frac{1}{2} \;\;{\mbox{and}} \;\; 2\lambda^C_3 + 2\lambda^C_2 -1 \neq 0\cr
1, &\displaystyle\lambda^B =\frac{1}{2}\;\;{\mbox{or}} \;\;2\lambda^C_3 + 2\lambda^C_2 -1 = 0
\end{array}
\right. .
\end{eqnarray}
{{}Eq.(\ref{ran}) shows that,} 
 if ${\mbox{ran}}\,\hat T=3$ (i.e., $\det \,\hat T\neq 0$), 
 then we have the complete information transfer from the first node  to the last one 
 (i.e., from $A$ to $B$). Consequently,  all parameters encoded into 
 $\rho^A(\varphi^A,0)$ are transfered to the subsystem $B$, so that $E^{AB}(t)=E^{AA}$. 
If ${\mbox{ran}}\,\hat T=1$, (i.e., either $\displaystyle\lambda^B=\frac{1}{2}$ or 
$2 \lambda^C_3 + 2\lambda^C_2 =1$)
then there is only one nonzero element in the 
matrix $J(\rho^B(\varphi^A,t))$ (the first order minor):
\begin{eqnarray}\label{ex1mn1}
M_1=-\frac{(2 \lambda^A_1-1) r^2(t) \sin(2\varphi_1)}{10(3+\sqrt{5})},
\end{eqnarray}
which is nonzero for  $\varphi_1\in G^A$. Thus, eq.(\ref{EAB}) yields $E^{AB}(t)=1$  if only 
$\displaystyle \lambda^A_1\neq \frac{1}{2}$. Otherwise, if $\displaystyle \lambda^A_1=\frac{1}{2}$, then  the rank of the matrix $J(\rho^B(\varphi^A,t))$ equals zero, and eq.(\ref{EAB}) yields  $E^{AB}(t)=0$.

All in all, 
\iffalse
we may write the following formula for  the informational correlation $E^{AB}(t)$:
\begin{eqnarray}\label{A1_t}
E^{AB}(t)=\left\{\begin{array}{ll}
2,& \displaystyle \lambda^A\neq \frac{1}{2},\; \lambda^B \neq \frac{1}{2},\; 2\lambda^C_3 + 2\lambda^C_2 -1\neq 0 \cr
1,&\displaystyle \lambda^A\neq \frac{1}{2} \;\; {\mbox{and}}\;\; \Big(\,\lambda^B=\frac{1}{2}, \; \;{\mbox{or }}\;
 2\lambda^C_3 + 2\lambda^C_2 -1 = 0\Big)
\cr  
0,& \displaystyle \lambda^A=\frac{1}{2} 
\end{array}\right..
\end{eqnarray}
In addition, for the normalized correlation $E^{AB}_{norm}(t)$ introduced  by eq.(\ref{norm}) we have ($\tilde D=2$):
\begin{eqnarray}\label{A1_tnorm}
E^{AB}_{norm}(t)=\left\{\begin{array}{ll}
1,&\displaystyle  \lambda^A\neq \frac{1}{2}, \,\lambda^B \neq \frac{1}{2},\; 2\lambda^C_3 + 2\lambda^C_2 -1\neq 0 \cr
\frac{1}{2},&\displaystyle \lambda^A\neq \frac{1}{2} \;\; {\mbox{and}}\;\; \Big(\,\lambda^B=\frac{1}{2}, \; \;{\mbox{or }}\;
 2\lambda^C_3 + 2\lambda^C_2 -1 = 0\Big)
 \cr  
0,&\displaystyle  \lambda^A=\frac{1}{2} 
\end{array}\right..
\end{eqnarray}
Eq.(\ref{A1_t}) allows us to conclude that the informational correlation  $E^{AB}$ is very sensitive to the multiplicity of the eigenvalues of the matrices $\rho^A(0)$ and $\rho^B(0)$.
\fi
{{}we obtain the values for $E^{AB}$ and $E^{AB}_{norm}$ collected 
in Table 1.}

%\multicolumn{3}{p{3.6cm}|}{mag. field of two currents, $b=1011.150$}&

\iffalse
\begin{table}{{}
\begin{tabular}{|l|l|l|l|}
\hline
 & ${\mbox{ran}} \hat T$ & $\lambda^A\neq 1/2$ & $\lambda^A=1/2$ \cr
 \hline
$E^{AA}$ & $\forall {\mbox{ran}} \hat T$&  2 & 0 \cr
$E^{AB}$ & 3 & 2 & 0 \cr
$E^{AB}$ & 1 & 1 & 0 \cr
\hline
\end{tabular}}
\caption{Parameter $E^{AA}$ and informational correlation  in dependence on eigenvalue $\lambda^A $ and rank of the matrix $\hat T$}
\end{table}
\fi
\begin{table}{{}
\begin{tabular}{|l|l|l|l|l|l|l|}
\hline
 & \multicolumn{2}{p{2cm}|}{$ \forall \;{\mbox{ran}} \hat T $ } & \multicolumn{2}{p{2cm}|}{$ {\mbox{ran}} \hat T =3 $ }& 
  \multicolumn{2}{p{2cm}|}{$ {\mbox{ran}} \hat T =1$  }  \cr
  \hline
 $\lambda^A$& $E^{AA}$ &$E^{AA}_{norm}$ & $E^{AB}$ &$E^{AB}_{norm}$ & $E^{AB}$ &$E^{AB}_{norm}$ \cr
  \hline
 $\lambda^A\neq 1/2$ &2&1&2&1&1&1/2\cr
 $\lambda^A=1/2$&0&0&0&0&0&0\cr
 \hline
%$E^{AA}$ & $\forall {\mbox{ran}} \hat T$&  2 & 0 \cr
%$E^{AB}$ & 3 & 2 & 0 \cr
%$E^{AB}$ & 1 & 1 & 0 \cr
\end{tabular}}
\caption{The parameter $E^{AA}$ ($E^{AA}_{norm}$) and the 
informational correlation $E^{AB}$ ($E^{AB}_{norm}$)
in dependence on the eigenvalue $\lambda^A $ and the rank of 
the matrix $\hat T$; $\tilde D^A=2$}
\end{table}

%%%%%%%%%%%%%%%%%
\subsubsection{Local initial  unitary transformations of the subsystems $C$ and $B$.}
\label{Section:loc}
The initial local unitary transformation of the subsystem $B$ has general form (\ref{UA}) 
with parameters $\beta_i$ instead of $\varphi_i$.   {{} 
This transformation} changes the  matrix $\hat T(t)$ given by eq.(\ref{hT}). Namely,  
factor $\cos(2\beta_1)$ appears  in the expressions for $a_1$ given by the first of 
eqs.(\ref{a1a2b}). Expression for $a_2$ (the second  of eqs.(\ref{a1a2b})) remains 
unchanged. {{}In addition,}  two more nonzero entries appear in the third row of $\hat T$. 
{{} All these modifications  change } 
the conditions in the rhs of  eqs.(\ref{ran}), {{}  which }
now reads:
\begin{eqnarray}\label{ranB}
&&
{\mbox{ran}} \,\hat T(t) =\left\{
\begin{array}{ll}
3 ,&\displaystyle \lambda^B \neq \frac{1}{2} , \;\; 2\lambda^C_3 + 2\lambda^C_2 -1 \neq 0,\;\;\beta_1\neq \frac{\pi}{4}
\cr
1, &\displaystyle\lambda^B =\frac{1}{2},\;\;{\mbox{or}} \;\;2\lambda^C_3 + 2\lambda^C_2 -1 = 0,\;\;{\mbox{or}}\;\;\beta_1 =  \frac{\pi}{4}
\end{array}
\right. .
%,\\
%\label{A1_tB}
%&&
%E^{AB}(t)=\left\{\begin{array}{ll}
%2,& \displaystyle \lambda^A\neq \frac{1}{2},\; \lambda^B \neq \frac{1}{2},\; 2\lambda^C_3 + 2\lambda^C_2 -1\neq 0,\;\;\beta_1\neq \frac{\pi}{4} \cr
%1,&\displaystyle \lambda^A\neq \frac{1}{2},\;\;{\mbox{and}} \;\;\Big(\lambda^B=\frac{1}{2}, \; \;{\mbox{or }}\;
%\; 2\lambda^C_3 + 2\lambda^C_2 -1 = 0,\;\;{\mbox{or}}\;\;\beta_1 =  \frac{\pi}{4} \Big)\cr  
%0,& \displaystyle \lambda^A=\frac{1}{2} 
%\end{array}\right..
\end{eqnarray}

Next, the initial local transformation of the subsystem $C$  changes formula (\ref{ran}) 
%,\ref{A1_t},\ref{A1_tnorm}) 
as well. To demonstrate this effect, we consider a  simple example. 
Let us take the following initial density matrix of the subsystem $C$ 
(instead of the diagonal initial state $\rho^C(0)$ 
(see eq.(\ref{N4rho01}))):
\begin{eqnarray}\label{CC}
\tilde \rho^C(0)=U^C \rho^C(0) (U^C)^+,\;\;
U^C=\left(
\begin{array}{cccc}
\cos \gamma & \sin\gamma&0&0\cr
-\sin\gamma & \cos \gamma & 0&0\cr
0&0&1&0\cr
0&0&0&1
\end{array}
\right),\;\;0<\gamma<2\pi.
\end{eqnarray} 
{{} 
The  local transformation $U^C$  effects on the matrix $\hat T $  
replacing  the expression  
$(2\lambda^C_3 + 2\lambda^C_2 -1)$    with
$(2\lambda^C_3+
 \lambda^C_2+\lambda^C_1 + (\lambda^C_2-\lambda^C_1)\cos(2 \gamma) -1)$ in eq.(\ref{ran}).}
 %,
 %\ref{A1_t},\ref{A1_tnorm}).

Results obtained in this subsection  mean that applying the local 
transformation we may handle the informational correlation up to a certain extent.
For instance, let the
subsystem $B$ be effected by the unitary transformation (outlined in 
the beginning of this subsection) 
with $\displaystyle \beta_1=\frac{\pi}{4}$.  {{} 
Then,  ${\mbox{ran}} \,\hat T$ decreases from $3$ to $1$. In turn, this
decreases  $E^{AB}$ from $2$ to $1$, 
as follows from  Table 1.}
Thus, this transformation may be used as the lock in some gates. Vice-versa,
 let $\displaystyle\lambda^A\neq \frac{1}{2}$ and the density matrix 
 of the subsystem $C$ be given by eq. (\ref{N4rho01}) with  $2\lambda^C_3 + 2\lambda^C_2 -1=0$. 
 Then, applying the local transformation $U^C$ with $\cos (2 \gamma)\neq 1$  {{} 
 (see eq.(\ref{CC})) we increase ${\mbox{ran}}\, \hat T$  
 from $1$ to $3$. In turn, this increases 
 $E^{AB}$ from $1$ to $2$, see Table 1. }

 %%%%%%%%%%%%%%%%%%%%
 \subsubsection{Non-reducible informational correlation} 
 \label{Section:nonredEX1}
In this example, the upper estimation (\ref{upper}) {{}for the non-reducible informational correlation}  yields $E^{AB;min}\le 1$. 
Remember that the evolution of the density matrix  is given by eq.(\ref{rhot}) with the initial density matrix (\ref{rho0ex1}).
 To calculate the non-reducible  informational correlation $E^{AB;min}$ {{}
 we use} the formulas of Sec.\ref{Section:nonred}. {{}Therewith}
 the density matrix $\rho^B(\varphi^A,t)$ {{} can be written as}:
  \begin{eqnarray}
 &&
 \rho^B(\varphi^A,t)=\left(\begin{array}{cc} \hat b(\varphi^A,t)
& \displaystyle
 \hat a(\varphi^A,t)  e^{-i\varphi_2}\cr\displaystyle
- \hat a(\varphi^A,t) e^{i\varphi_2} & \displaystyle 1-\hat b(\varphi^A,t)
\end{array}\right),\\\nonumber
&&\hat b(\varphi^A,t)= b(t) +\frac{a_2(t)}{2} ( (2\lambda^A_1-1) \cos(2\varphi_1)+1),
\;\;\;
\hat a(\varphi^A,t)=-\frac{i a_{1}(t)}{2} (2 \lambda^A_1-1) \sin(2\varphi_1),
  \end{eqnarray}
  where $a_1$, $a_2$ and $b$  are given in eq.(\ref{a1a2b}).
  To calculate $E^{AB;min}$ using  eq.(\ref{JH}) we, first of all, shall write  
  the  characteristic equation for the matrix $\rho^B(\varphi^A,t)$:
\begin{eqnarray}
\lambda^2-\lambda + \det \,\rho^B(\varphi^A,t) =0.
\end{eqnarray}
It is obvious that $\det (\rho^B(\varphi^A,t))$ depends  on $\varphi_1$ and does not depend on $\varphi_2$. 
Consequently, representation (\ref{HH}) of $H$ becomes  scalar:
\begin{eqnarray}
H(\varphi_1,t)&=&\frac{\partial\det \,\rho^B(t)}{\partial\varphi_1} = 
a_2(t) (2\lambda^A_1-1) \sin(2\varphi_1) \Big( 2 b(t) +a_2(t) -1 + \\\nonumber
&& (2\lambda^A_1-1) (a_2(t) -(2\lambda^B_1-1)^2 (2\lambda^C_3 + 2\lambda^C_2-1)^2)\cos(2\varphi_1)\Big).
\end{eqnarray}
Thus eq.(\ref{JH}) yields
\begin{eqnarray}\label{EABmin}
&&
E^{AB;min}(\varphi_1,t)={\mbox{ran}}\, H(\varphi_1,t) = \left\{\begin{array}{ll}
1, &H(\varphi_1,t) \neq 0\cr
0, &H(\varphi_1,t)= 0
\end{array}\right.,\\\nonumber
&&
E^{AB;min}_{norm}(\varphi_1,t)= \left\{\begin{array}{ll}
\displaystyle\frac{1}{2}, &H(\varphi_1,t) \neq 0\cr
0, &H(\varphi_1,t)= 0
\end{array}\right..
\end{eqnarray}
All zeros  of the function $H(\varphi_1,t)$ are defined by the following formulas (remember, that $\displaystyle 
0 <  \varphi_1 <  \frac{\pi}{2}$ as indicated in eq.(\ref{int})):
\begin{eqnarray}\label{z1}
&&r=0,\\\label{z2}
&&\lambda^A_1=\frac{1}{2},\\\label{z3}
&&\sin(2\varphi_1)=0\;\;\Rightarrow \;\;
\varphi_1 = 0,\frac{\pi}{2},\\\label{z4}
&&\cos 2\varphi_1 = m(t),\;\;\Rightarrow \;\;\\\label{z40}
&& \varphi_1 =\frac{1}{2}\arccos\,m(t),\\\nonumber
&&\hspace{1cm}
m(t)=-\frac{2 b(t) +a_2(t) -1 }{(2\lambda^A_1-1) ( a_2(t) -
 (2\lambda^B_1-1)^2  (2\lambda^C_3 + 2\lambda^C_2 -1)^2 )}.
\end{eqnarray}
Analyzing eqs.(\ref{z1}-\ref{z40}) we see,
first of all,  that $H(\varphi_1,t)$ is identical to zero for all $\varphi_1$ at 
the time instants satisfying  condition (\ref{z1}). {{}But} these instants are excluded by condition (\ref{rneq0}). 
We obviously shall not use the initial matrix $\rho^A(0)$ with equal eigenvalues, that follows from eq.(\ref{z2}).
Next, eq.(\ref{z3}) means that the boundary values $0$ and $\displaystyle\frac{\pi}{2}$ 
of the parameter $\varphi_1$ may not be uniquely transfered at any time instant. However, the boundary is not involved into our consideration. 
Finally, $E^{AB;min}=0$ at instants satisfying  eq.(\ref{z4}).
Thus, if at some instant $t_1$ (provided that $\varphi_1\in G^A$,  
$\displaystyle \lambda^A\neq \frac{1}{2}$ and $r(t_1)\neq 0$) we obtain  
$H(\varphi^A,t_1)=0$,  this means that  the  value of the 
parameter $\varphi_1$ defined by eq.(\ref{z40}) is transfered.  
To avoid the zero value of $H$,   we shall consider such time instants which do not satisfy condition (\ref{z4}) for all $\varphi_1$ from the interval $\displaystyle 0 < \varphi_1 < \frac{\pi}{2}$, i.e., 
\begin{eqnarray}\label{mpos}
|m(t)|>1.
\end{eqnarray}
This may be done for any particular set of fixed eigenvalues of the initial density matrices $\rho^A(0)$, $\rho^C(0)$ and  $\rho^B(0)$.
For instance, let us define the suitable time  interval for the following set of   eigenvalues:
\begin{eqnarray}\label{evconst}
\lambda^A_1=\lambda^B_1=\frac{3}{4}, \;\;\lambda^C_1=\lambda^C_2=\lambda^C_3=\frac{1}{4}.
\end{eqnarray}
The first positive root of equation  $|m(t)|=1$ is $t=2.726$. Consequently, the first positive   time interval where  condition (\ref{mpos}) is satisfied (and consequently condition (\ref{z4}) fails)  is the following one:
\begin{eqnarray}\label{int1}
0<t<2.726.
\end{eqnarray}
Interval (\ref{int1})  is suitable for the  detection of the  parameter $\varphi_1$.
In this case $g=G^A$ and we conclude that 
$E^{AB;min}(G^A,t)=1$  for the eigenvalues (\ref{evconst}) if $t$ is inside of the interval (\ref{int1}).

We see that  the  parameter $\varphi_1$ is encoded into the eigenvalue of the matrix $\rho^B(\varphi^A,t)$. It is obvious that $\varphi_1$ is also encoded into the matrix of eigenvectors of  $\rho^B(\varphi^A,t)$, because otherwise two matrices with different values of $\varphi_1$ would have the same complete set of independent eigenvectors and consequently they would commute which is not true (this might be checked directly in our example). Therefore, the
parameter $\varphi_1$ is most reliable one  since it  might be detected in the  either
 eigenvalue or  eigenvectors  
of the  density matrix $\rho^B$. Meanwhile, the parameter $\varphi_2$  may  be transfered only by the 
eigenvectors  of the density matrix  $\rho^B(\varphi^A,t)$ and may be removed by the local unitary transformation 
of the subsystem $B$ at instant $t$. Again, this result is valid provided that condition (\ref{rneq0}) is satisfied. It 
might be shown that the   local transformations performed on the either subsystem $C$ or $B$  may insert 
the parameter $\varphi_2$ into the determinant $\det\,\rho^B(\varphi^A,t)$ = $H(\varphi^A,t)$, which 
appears as a condition in the rhs of eq.(\ref{EABmin}). Both parameters $\varphi_1$ and $\varphi_2$ may be used on the equal foot in this case. 

%%%%%%%%%%%%%%%%%%%
\subsection{Two-node subsystems $A$ and $B$}
Now we consider the informational correlation between two-node subsystems 
$A$ and $B$ of the four-node homogeneous  chain with  XY Hamiltonian (\ref{H}). 
The subsystem $A$ is represented by the first and the second nodes, while the subsystem 
$B$ is represented by the third and the fourth nodes. Thus  $N^A=N^B=4$, $N^C=0$. 
The group of the local  unitary transformations of the subsystem $A$ {{}
is $SU(4)$.}
We consider the initial density matrix in the form (\ref{inden}) without the subsystem $C$:
\begin{eqnarray}\label{rho0ex2}
\rho(0) =\rho^A(0) \times \rho^B(0),
\end{eqnarray}
where $\rho^A(0)$ and $\rho^B(0)$ are the diagonal matrices:
\begin{eqnarray}\label{lamrel}
&&
\rho^A(0)={\mbox{diag}}(\lambda^A_1,\lambda^A_2,\lambda^A_3,\lambda^A_4),\;\;\;
\lambda^A_1\ge\lambda^A_2\ge \lambda^A_3\ge\lambda^A_4,\;\;\sum_{i=1}^4\lambda^A_i=1,\\\nonumber
&&
\rho^B(0)={\mbox{diag}}(\lambda^B_1,\lambda^B_2,\lambda^B_3,\lambda^B_4),\;\;\;
\lambda^B_1\ge\lambda^B_2\ge \lambda^B_3\ge\lambda^B_4,\;\;\sum_{i=1}^4\lambda^B_i=1.
\end{eqnarray}

%%%%%%%%%%%%%
\subsubsection{Number of parameters encoded into the subsystem $A$, $E^{AA}$}
\label{Section:EAAn2}
The group $SU(4)$  is the 15-parametric one.
However, 
considering the transformation of the diagonal matrix
$\rho^A(0)$, we deal with the 12-parametric representation, $\tilde D^A=12$ 
(see eq.(\ref{DA}) and ref.\cite{TBS}):
\begin{eqnarray}\label{UrhoU}
&&
\rho^A(\varphi^A,0)= U^A(\varphi^A) \Lambda^A (U^A(\varphi^A))^+=\tilde U^A(\varphi^A) \Lambda^A (\tilde U^A(\varphi^A))^+,\\\label{UrhoUp}
&&
\tilde U^A(\varphi^A)=e^{i\gamma_3 \varphi_1}e^{i\gamma_2 \varphi_2}e^{i\gamma_3 \varphi_3}e^{i\gamma_5 \varphi_4}
e^{i\gamma_3 \varphi_5}e^{i\gamma_{10} \varphi_6}e^{i\gamma_3 \varphi_7}e^{i\gamma_2 \varphi_8}e^{i\gamma_3 \varphi_9}
e^{i\gamma_5 \varphi_{10}}e^{i\gamma_{3} \varphi_{11}}e^{i\gamma_2 \varphi_{12}}.
\end{eqnarray}
Here the ranges of the parameters $\varphi_i$ are following \cite{TBS}
\begin{eqnarray}\label{phi_reg}
&&
0\le \varphi_1,\varphi_3, \varphi_5,\varphi_7, \varphi_9,\varphi_{11} \le \pi,\;\;\;%\\\nonumber
%&&
0\le \varphi_2,\varphi_4,\varphi_6,\varphi_8,\varphi_{10},\varphi_{12} \le \frac{\pi}{2}.
\end{eqnarray}
The explicit matrix representation  of $\gamma_i$ is given in the Appendix A, Sec.\ref{Section:A} \cite{TBS,GM}. One should note that the expression (\ref{UrhoUp})
for $\tilde U^A$ holds if all eigenvalues $\lambda^A_i$ are different. In this case $E^{AA}=12$.
However, not all 12  parameters may be encoded into the density matrix $\rho^A(\varphi^A,0)$ if some of eigenvalues $\lambda^A_i$ coincide.  
For instance, if $\lambda^A_1=\lambda^A_2$, then $\gamma_2$ and  
$\gamma_3$ commute with $\rho^A(0)$. {{} Then}, eqs.(\ref{UrhoU},\ref{UrhoUp}) reduce to
\begin{eqnarray}\label{UrhoU2}
&&
\rho^A(\varphi^A,0)=\tilde U^A_2(\varphi^A) \Lambda^A (\tilde U^A_2(\varphi^A))^+,\\\nonumber
&&
\tilde U^A_2(\varphi^A)=e^{i\gamma_3 \varphi_1}e^{i\gamma_2 \varphi_2}e^{i\gamma_3 \varphi_3}e^{i\gamma_5 \varphi_4}
e^{i\gamma_3 \varphi_5}e^{i\gamma_{10} \varphi_6}e^{i\gamma_3 \varphi_7}e^{i\gamma_2 \varphi_8}e^{i\gamma_3 \varphi_9}
e^{i\gamma_5 \varphi_{10}},
\end{eqnarray}
where $\tilde U^A_2$  possesses 10 parameters $\varphi_i$, $i=1,\dots,10$. {{}
Thus}, only 10 parameters may 
be encoded into the density matrix $\rho^A(\varphi^A)$, i.e., $E^{AA}= 10$, which agrees with eq.(\ref{tDK}).

Next, let  $\lambda^A_1=\lambda^A_2$ and $\lambda^A_3=\lambda^A_4$,  but $\lambda^A_1 \neq \lambda^A_3$. Then,  eq.(\ref{tDK}) yields 
 $E^{AA}= 8$.

If $\lambda^A_1=\lambda^A_2=\lambda^A_3$, then $\gamma_2$, $\gamma_3$ and $\gamma_5$ commute with $\rho^A(0)$, so that eq.(\ref{UrhoU2}) reduces to 
\begin{eqnarray}\label{UrhoU3}
&&
\rho^A(\varphi^A,0)=\tilde U^A_3(\varphi^A) \Lambda^A (\tilde U^A_3(\varphi^A))^+,\\\nonumber
&&
\tilde U^A_3(\varphi^A)=e^{i\gamma_3 \varphi_1}e^{i\gamma_2 \varphi_2}e^{i\gamma_3 \varphi_3}e^{i\gamma_5 \varphi_4}
e^{i\gamma_3 \varphi_5}e^{i\gamma_{10} \varphi_6},
\end{eqnarray}
where $\tilde U^A_3$  possesses 6 parameters $\varphi_i$, $i=1,\dots,6$. Consequently only 6 parameters may be encoded into the density matrix $\rho^A(\varphi^A)$, i.e., $E^{AA}= 6$. This also agrees with eq.(\ref{tDK}).

Finally, if  $\lambda^A_1=\lambda^A_2=\lambda^A_3=\lambda^A_4=1/4$, then $\rho^A(\varphi^A,0)$ is proportional to the $4\times 4$ identity matrix  $I_4$,
\begin{eqnarray}\label{UrhoU4}
&&
\rho^A(\varphi^A,0)= \frac{1}{4} I_4.
\end{eqnarray}
No parameters may be encoded into such $\rho^A(\varphi^A,0)$, i.e., $E^{AA}=0$.

{{} The calculated  values of $E^{AA}$ (and $E^{AA}_{norm}$) 
depend on $\lambda^A_i$, which is shown in   Table 2.}
\iffalse
\begin{eqnarray}\label{A1_02}
E^{AA}=\left\{\begin{array}{ll}
12,&\lambda^A_1\neq \lambda^A_2 \neq \lambda^A_3\neq \lambda^A_4\cr
10,& \lambda^A_1=\lambda^A_2, \lambda^A_1\neq \lambda^A_3\neq \lambda^A_4\cr
8,& \lambda^A_1=\lambda^A_2, \lambda^A_4= \lambda^A_3,\lambda^A_1\neq \lambda^A_3\cr
6,& \lambda^A_1=\lambda^A_2=\lambda^A_3\neq \lambda^A_4\cr
0,&\displaystyle\lambda^A_1=\lambda^A_2=\lambda^A_3= \lambda^A_4=\frac{1}{4}
\end{array}\right..
\end{eqnarray}
In addition, for the normalized parameter $E^{AA}_{norm}$ given by eq.(\ref{normA}), we find
\begin{eqnarray}\label{TAB1norm}
E^{AA}_{norm}=\left\{\begin{array}{ll}
1,&\lambda^A_1\neq \lambda^A_2 \neq \lambda^A_3\neq \lambda^A_4\cr
\displaystyle\frac{5}{6},& \lambda^A_1=\lambda^A_2, \lambda^A_1\neq \lambda^A_3\neq \lambda^A_4\cr
\displaystyle\frac{2}{3},& \lambda^A_1=\lambda^A_2, \lambda^A_4= \lambda^A_3,\lambda^A_1\neq \lambda^A_3\cr
\displaystyle\frac{1}{2},& \lambda^A_1=\lambda^A_2=\lambda^A_3\neq \lambda^A_4\cr
0,&\displaystyle\lambda^A_1=\lambda^A_2=\lambda^A_3= \lambda^A_4=\frac{1}{4}
\end{array}\right..
\end{eqnarray}
\fi
Remark that the same {{} values} of $E^{AA}$ may be obtained 
calculating the matrix  $J^A(\varphi^A,0)$  and using eq.(\ref{EAA}). Therewith 
the matrix $\hat X(\rho^A(\varphi^A,0))$ may be constructed by its  
definition (\ref{XYZ},\ref{XT}). {{} But} we do not represents
its  explicit form here, because it is too complicated. 

%%%%%%%%%%%%%%%
\subsubsection{Relation between the  rank of $\hat T$ and the informational correlation $E^{AB}$}
\label{Section:ranT2}
Now we apply the  transformation $V(t)\in SU(16)$ defined by eq.(\ref{Wt})
with Hamiltonian (\ref{H}) and find the density matrix $\rho(\varphi^A,t)$ of 
the system using eq.(\ref{rhot}),
where
\begin{eqnarray}
\rho(\varphi^A,0)= (\tilde U^A(\varphi^A) \times I_4) \rho(0) 
( ( \tilde U^A(\varphi^A))^+ \times I_4) ,
\end{eqnarray}
and the initial density matrix  $\rho(0)$ is given by eq.(\ref{rho0ex2}).
{{}Next,} we calculate the matrix $\hat T$ using 
eqs.(\ref{T},\ref{hTT1},\ref{hTT3},\ref{hTT4}). The explicit form of this matrix 
is very complicated and it is not represented
in this paper. Below  we study the informational correlation  $E^{AB}$ 
for different  ranks of the matrix $\hat T$. 

%%%%%%%%%%%%%%%%
\paragraph{Complete information transfer, ${\mbox{ran}}\,\hat T =15$ (or $\det\,\hat T\neq 0$).}
The direct calculations show that the condition for the complete information transfer reads
\begin{eqnarray}\label{M15}
\det \hat T =-\frac{(2 \lambda^B_2 + 2 \lambda^B_3-1)^8}{10^{16}}\Big(
\cos\frac{\sqrt{5} t}{2} -5 \cos\frac{t}{2}  +4 
\Big)^{16}\neq 0.
\end{eqnarray}
In this case all parameters $E^{AA}$ will be transfered into the subsystem $B$, i.e., $E^{AB}(t)=
E^{AA}$ and $E^{AB}_{norm}(t)=
E^{AA}_{norm}$, see Table 2. %eqs.(\ref{A1_02},\ref{TAB1norm}). 
Therewith ${\mbox{ran}}\,\hat T =15$. 
Hereafter we consider  such instants $t$ that 
the second factor in eq.(\ref{M15}) is non-zero, i.e.,
\begin{eqnarray}\label{zerot}
\cos\frac{\sqrt{5} t}{2} -5 \cos\frac{t}{2}  +4  \neq 0.
\end{eqnarray}
The first positive root of the expression in the lhs of  (\ref{zerot}) is $t=11.909$. 
Thus, the suitable time interval for the parameter detection might be 
\begin{eqnarray}\label{int11}
0<t<11.909.
\end{eqnarray}

%%%%%%%%%%%%%%%
\paragraph{Partial information transfer, ${\mbox{ran}}\,\hat T =11$.}
Next, let $\lambda^B_i$ be such that 
$\det \hat T =0$, i.e.,
\begin{eqnarray}\label{lamB1}
\lambda^B_3=\frac{1}{2} -\lambda^B_2.
\end{eqnarray}
Then we have to calculate the rank of the matrix $\hat T$. It might be readily checked that   the 11th order minors of $\hat T$ are nonzero and they may be represented by the following formula:
\begin{eqnarray}\label{M11}
M_{11}&=&\pm
\frac{(\lambda^B_2 (2\lambda^B_2-1) -\lambda^B_1 (2\lambda^B_1-1) )^2}{4\times 10^{14}}\Big(
\cos\frac{\sqrt{5} t}{2} -5 \cos\frac{t}{2}  +4 
\Big)^{10}\times \\\nonumber
&&\Big(
\cos\frac{\sqrt{5} t}{2} +5 \cos\frac{t}{2}  +4 
\Big)^{4}.
\end{eqnarray}
Thus, ${\mbox{ran}}\,\hat T =11$. 
\iffalse
\begin{eqnarray}\label{TAB2}
E^{AB}(t)=\left\{\begin{array}{ll}
11,&\lambda^A_1\neq \lambda^A_2 \neq \lambda^A_3\neq \lambda^A_4\cr
10,& \lambda^A_1=\lambda^A_2, \lambda^A_1\neq \lambda^A_3\neq \lambda^A_4\cr
8,& \lambda^A_1=\lambda^A_2, \lambda^A_4= \lambda^A_3,\lambda^A_1\neq \lambda^A_3\cr
6,& \lambda^A_1=\lambda^A_2=\lambda^A_3\neq \lambda^A_4\cr
0,&\displaystyle\lambda^A_1=\lambda^A_2=\lambda^A_3= \lambda^A_4=\frac{1}{4}
\end{array}\right..
\end{eqnarray}
In addition, for the normalized informational correlation $E^{AB}_{norm}(t)$ given by eq.(\ref{norm}), we obtain
{{}(remember that $\tilde D^A=12$)}
\begin{eqnarray}\label{TAB2norm}
E^{AB}_{norm}(t)=\left\{\begin{array}{ll}
\displaystyle \frac{11}{12},&\lambda^A_1\neq \lambda^A_2 \neq \lambda^A_3\neq \lambda^A_4\cr
\displaystyle \frac{5}{6},& \lambda^A_1=\lambda^A_2, \lambda^A_1\neq \lambda^A_3\neq \lambda^A_4\cr
\displaystyle \frac{2}{3},& \lambda^A_1=\lambda^A_2, \lambda^A_4= \lambda^A_3,\lambda^A_1\neq \lambda^A_3\cr
\displaystyle \frac{1}{2},& \lambda^A_1=\lambda^A_2=\lambda^A_3\neq \lambda^A_4\cr
0,&\displaystyle\lambda^A_1=\lambda^A_2=\lambda^A_3= \lambda^A_4=\frac{1}{4}
\end{array}\right..
\end{eqnarray}
\fi
%
{{} Calculating $E^{AB}$ }
we shall consider such instant $t$ that two last factors in the formula (\ref{M11}) are nonzero. In other  words,  along with condition (\ref{zerot}), the following condition must be  satisfied 
\begin{eqnarray}\label{zerot2}
\cos\frac{\sqrt{5} t}{2} +5 \cos\frac{t}{2}  +4  \neq 0.
\end{eqnarray}
The first positive root of the  expression in the lhs of condition (\ref{zerot2}) is $t=5.952$ 
which {{}reduces} interval (\ref{int11}) to $0<t<5.952$.

The informational correlation  $E^{AB}$ can be calculated by  formula (\ref{EAB}).
{{} It 
depends  on $\lambda_i$, which is shown in Table 2. 
$E^{AB}_{norm}(t)$ is represented in Table 2 as well
($\tilde D^A=12$).}

%%%%%%%%%%%%%%%
\paragraph{Partial information transfer, ${\mbox{ran}}\,\hat T =9$.}

Next, we consider such $\lambda^B_i$ that $M_{11}=0$. This happens in one of  two following cases:
\begin{eqnarray}\label{lam1}
&&
\lambda^B_2=\lambda^B_1, \;\;{\mbox{or}}\\
\label{lam2}
&&
\lambda^B_2=\frac{1}{2}-\lambda^B_1.
\end{eqnarray}
In the first case, eq.(\ref{lam1}), we have two following pairs of equal eigenvalues:
\begin{eqnarray}\label{lam12}
\lambda^B_4=\lambda^B_3=\frac{1}{2}-\lambda^B_1,\;\;\lambda^B_2=\lambda^B_1.
\end{eqnarray}
In the second case, eq.(\ref{lam2}), we have two other pairs of equal eigenvalues:
\begin{eqnarray}\label{extra}
\lambda^B_3=\lambda^B_1=\frac{1}{2}-\lambda^B_2,\;\;\lambda^B_4=\lambda^B_2.
\end{eqnarray}
In both cases\footnote{In principle, the case (\ref{extra}) might be disregarded because it does not meet the relations among the eigenvalues given in eq.(\ref{lamrel}).},  
the 9th order minors $M_{9i}$ of $\hat T$ are nonzero and all of them are collected in the following formula:
\begin{eqnarray}\label{M9}
M_{9i}= -\frac{\Big(
\cos\frac{\sqrt{5} t}{2} -5 \cos\frac{t}{2}  +4 
\Big)^{8}}{10^8} (4 \lambda^B_1-1) m_i(t),\;\;i=1,\dots,16,
\end{eqnarray}
where $m_i(t)$ are  some explicit functions of $t$
independent on $\lambda^B_i$, but they depend on whether eq.(\ref{lam1}) or (\ref{lam2}) is considered (we do not represent the expressions for these functions). The analysis shows that 
 $\displaystyle\sum_{i=1}^{16} |m_i(t)| \neq 0$ for all $t$ in both cases (\ref{lam1}) and (\ref{lam2}). 
 Thus, ${\mbox{ran}}\,\hat T =9$  provided  that condition  (\ref{zerot}) is satisfied  and  $\lambda^B_1\neq \frac{1}{4}$. 

{{} The informational correlation  $E^{AB}$ calculated by 
 formula (\ref{EAB}) depends on $\lambda^A_i$, which is shown in Table 2, where 
the appropriate values of $E^{AB}_{norm}$ are represented as well.} 
\iffalse
\begin{eqnarray}\label{TAB3}
E^{AB}(t)=\left\{\begin{array}{ll}
9,&\lambda^A_1\neq \lambda^A_2 \neq \lambda^A_3\neq \lambda^A_4\cr
9,& \lambda^A_1=\lambda^A_2, \lambda^A_1\neq \lambda^A_3\neq \lambda^A_4\cr
8,& \lambda^A_1=\lambda^A_2, \lambda^A_4= \lambda^A_3,\lambda^A_1\neq \lambda^A_3\cr
6,& \lambda^A_1=\lambda^A_2=\lambda^A_3\neq \lambda^A_4\cr
0,&\displaystyle\lambda^A_1=\lambda^A_2=\lambda^A_3= \lambda^A_4=\frac{1}{4}
\end{array}\right..
\end{eqnarray}
In addition, for the normalized informational correlation $E^{AB}_{norm}(t)$ (see eq.(\ref{norm})), we obtain
\begin{eqnarray}\label{TAB3norm}
E^{AB}_{norm}(t)=\left\{\begin{array}{ll}
\displaystyle\frac{3}{4},&\lambda^A_1\neq \lambda^A_2 \neq \lambda^A_3\neq \lambda^A_4\cr
\displaystyle\frac{3}{4},& \lambda^A_1=\lambda^A_2, \lambda^A_1\neq \lambda^A_3\neq \lambda^A_4\cr
\displaystyle\frac{2}{3},& \lambda^A_1=\lambda^A_2, \lambda^A_4= \lambda^A_3,\lambda^A_1\neq \lambda^A_3\cr
\displaystyle\frac{1}{2},& \lambda^A_1=\lambda^A_2=\lambda^A_3\neq \lambda^A_4\cr
0,&\displaystyle\lambda^A_1=\lambda^A_2=\lambda^A_3= \lambda^A_4=\frac{1}{4}
\end{array}\right..
\end{eqnarray}
\fi

%%%%%%%%%%%%%%%
\paragraph{Partial information transfer, ${\mbox{ran}}\,\hat T =7$.}
Finally, let $\lambda^B_1=1/4$, which means that all $\lambda^B_i$ are the same and equal
$1/4$. Then,  the 7th order minor of $\hat T$ is nonzero:
\begin{eqnarray}\label{M7}
M_7= -\frac{\Big(
\cos\frac{\sqrt{5} t}{2} -5 \cos\frac{t}{2}  +4 
\Big)^{8}}{10^8} .
\end{eqnarray}
Thus, ${\mbox{ran}}\,\hat T =7$.
{{}Again, the informational correlation  $E^{AB}$ calculated by  formula (\ref{EAB}) 
depends on $\lambda^A_i$, which is shown in Table 2, where 
the values of $E^{AB}_{norm}$ are represented as well.}
\iffalse
\begin{eqnarray}\label{TAB4}
E^{AB}(t)=\left\{\begin{array}{ll}
7,&\lambda^A_1\neq \lambda^A_2 \neq \lambda^A_3\neq \lambda^A_4\cr
7,& \lambda^A_1=\lambda^A_2, \lambda^A_1\neq \lambda^A_3\neq \lambda^A_4\cr
6& \lambda^A_1=\lambda^A_2, \lambda^A_4= \lambda^A_3,\lambda^A_1\neq \lambda^A_3\cr
5,& \lambda^A_1=\lambda^A_2=\lambda^A_3\neq \lambda^A_4\cr
0,&\displaystyle\lambda^A_1=\lambda^A_2=\lambda^A_3= \lambda^A_4=\frac{1}{4}
\end{array}\right..
\end{eqnarray}
In addition, for the normalized informational correlation $E^{AB}_{norm}(t)$ given by eq.(\ref{norm}), we obtain
\begin{eqnarray}\label{TAB4norm}
E^{AB}_{norm}(t)=\left\{\begin{array}{ll}
\displaystyle\frac{7}{12},&\lambda^A_1\neq \lambda^A_2 \neq \lambda^A_3\neq \lambda^A_4\cr
\displaystyle\frac{7}{12},& \lambda^A_1=\lambda^A_2, \lambda^A_1\neq \lambda^A_3\neq \lambda^A_4\cr
\displaystyle\frac{1}{2},& \lambda^A_1=\lambda^A_2, \lambda^A_4= \lambda^A_3,\lambda^A_1\neq \lambda^A_3\cr
\displaystyle\frac{5}{12},& \lambda^A_1=\lambda^A_2=\lambda^A_3\neq \lambda^A_4\cr
0,&\displaystyle\lambda^A_1=\lambda^A_2=\lambda^A_3= \lambda^A_4=\frac{1}{4}
\end{array}\right..
\end{eqnarray}
\fi

Similar to Sec.\ref{Section:ranT}, from the analysis of 
Table 2, we conclude that  the informational correlation $E^{AB}$ is very  sensitive to the multiplicity of the eigenvalues of the density matrices $\rho^A(0)$ and  $\rho^B(0)$.

We emphasize that the represented analysis is valid at the time instants   
satisfying conditions (\ref{zerot}) and, generally speaking, (\ref{zerot2}).

%{{}We collect the basic results of Secs.\ref{Section:EAAn2} and \ref{Section:ranT2}
%in Table 2.}

\begin{table}{{}
\begin{tabular}{|p{2.5cm}|p{1cm}|p{1cm}|p{1cm}|p{1cm}|p{1cm}|p{1cm}|p{1cm}|p{1cm}|p{1cm}|p{1cm}|}
\hline
 & \multicolumn{2}{p{2cm}|}{$ \forall \;{\mbox{ran}} \hat T $ } & \multicolumn{2}{p{2cm}|}{$ {\mbox{ran}} \hat T =15 $ }& 
  \multicolumn{2}{p{2cm}|}{$ {\mbox{ran}} \hat T =11$  } & 
  \multicolumn{2}{p{2cm}|}{$ {\mbox{ran}} \hat T =9$  } & 
  \multicolumn{2}{p{2cm}|}{$ {\mbox{ran}} \hat T =7$  }  \cr
  \hline
 $\lambda^A_1,\lambda^A_2,\lambda^A_3,\lambda^A_4 $& $E^{AA}$ &$E^{AA}_{norm}$ & $E^{AB}$ &$E^{AB}_{norm}$ & $E^{AB}$ &$E^{AB}_{norm}$  & $E^{AB}$ &
 $E^{AB}_{norm}$ & $E^{AB}$ &$E^{AB}_{norm}$\cr
  \hline
  $\lambda^A_1\neq \lambda^A_2\neq$ $ \lambda^A_3\neq \lambda^A_4$ &12&1&12&1    &11&11/12  &9&3/4  &7&7/12\cr\hline
$\lambda^A_1= \lambda^A_2\neq$ $ \lambda^A_3\neq \lambda^A_4$ &10&5/6&10&5/6     &10&5/6  &9&3/4 &7&7/12\cr\hline
$\lambda^A_1=\lambda^A_2$, $\lambda^A_3= \lambda^A_4$, $\lambda_1\neq  \lambda^A_3$ &8&2/3&8&2/3   &8&2/3  &8&2/3 &6&1/2\cr\hline
  $\lambda^A_1=\lambda^A_2=\lambda^A_3\neq  \lambda^A_4$ &6&1/2&6&1/2      &6&1/2   &6&1/2   &5&5/12\cr\hline
  $\lambda^A_1=\lambda^A_2= \lambda^A_3=$
  $\lambda^A_4=\frac{1}{4}$&0&0&0&0 &0&0  &0&0  &0&0  \cr
 \hline
\end{tabular}}
\caption{The parameter $E^{AA}$ ($E^{AA}_{norm}$) and the informational correlation  
$E^{AB}$ ($E^{AB}_{norm}$)
in dependence on the eigenvalues $\lambda^A_i $ ($i=1,2,3,4$)
and the rank of the matrix $\hat T$; $\tilde D^A=12$.}
\end{table}

\paragraph{ Local unitary transformations of the subsystem $B$.}  Similar to the example of 
Sec.\ref{Section:loc}, the local unitary transformations of the subsystem $B$ allow one to handle the informational correlation up to a certain extent. Having the initial density matrix $\rho^B(0)$ 
with the fixed eigenvalues, we may apply the local unitary transformation with the purpose to either increase or decrease the number of parameters transfered from the subsystem $A$ to the subsystem $B$. The detailed study of this problem is beyond the scope of this paper.

\subsubsection{Non-reducible informational correlation}
\label{Section:nonredEX2}
In this example, the upper estimation (\ref{upper}) yields 
$E^{AB;min}\le 3$. We may not carry out all  calculations of 
Sec.\ref{Section:nonred} analytically. Therefore we do not 
study the non-reducible correlation in the full extend. 
To illustrate some properties of the non-reducible correlations, 
we consider two particular examples.

\paragraph{ Example 1: ${\mbox{ran}} \,\hat T=7$,  $E^{AB}=5$ in Table 2.}
In {{} accordance with Table 2,}  $E^{AA}=6$, i.e., 
there are six parameters in the set $\varphi^A$. 
To calculate  the non-reducible informational correlation by formula (\ref{JH}),
we construct the $3\times 6$ matrix $H$ (\ref{HH}). It may be readily shown that  
the third order minors of the  matrix $H$ are zeros. The second order minors  are 
non-zeros  so that $E^{AB;min}(\varphi^A,t)= 2$. The analysis of these minors shows 
that the following pairs of parameters might be transfered from the subsystem $A$ 
to the subsystem $B$:
\begin{eqnarray}\label{pairs}
&&
(\varphi_1, \;\;\varphi_2), \;\;\;\;
(\varphi_1, \;\;\varphi_4), \;\;\;\;
(\varphi_1, \;\;\varphi_6), \;\;\;\;
(\varphi_2, \;\;\varphi_3), \;\;\;\;
(\varphi_2, \;\;\varphi_4), \\\nonumber
&&
(\varphi_2, \;\;\varphi_6), \;\;\;\;
(\varphi_3, \;\;\varphi_4), \;\;\;\;
(\varphi_3, \;\;\varphi_6), \;\;\;\;
(\varphi_4, \;\;\varphi_6).
\end{eqnarray}
It is remarkable that the parameter $\varphi_5$ does not appear in the above list (\ref{pairs}).

However, the second order minors {{} under consideration} depend on $\varphi^A$ and $t$.  
{{} Consequently,} there might be such $\varphi^A$ and $t$ that the second order minors equal to zero.
 {{} This} decreases $E^{AB;min}$. 
Thus, as the next step, we must define such  region $g\subset G^A$ and such time intervals that at least one of  
pairs (\ref{pairs}) with values from $g$ may be detected in the subsystem $B$ at any instant during these time intervals (only in that case $E^{AB;min}= 2$).

We do not consider this problem in the full extend. Instead of this, we study the problem of informational correlation between the subsystems $A$ and $B$ by means of a particular pair of parameters  from the list (\ref{pairs}).  For instance, let us consider the pair $(\varphi_2,\varphi_6)$. In other words, we restrict the set $\varphi^A$ to $\hat \varphi^A=\{\varphi_2,\varphi_6\}$ and  $G^A$ to $\hat G^A \in G^A$ defined as follows:
\begin{eqnarray}\label{Gex1}
\displaystyle \hat G^A: 0 <\varphi_2,\varphi_6<\frac{\pi}{2}.
\end{eqnarray}
This  {{}restriction} of the whole set of 6 parameters  in the unitary transformation of the subsystem $A$ 
($E^{AA} = 6$ in our case)  is justified by the fact that not all 
parameters of the local unitary transformation might be equivalent for a particular
quantum information problem. In this example we assume that namely parameters $\varphi_2$ and $\varphi_6$ must be used  
  {{} in a quantum process}. 

\begin{figure*}
   \epsfig{file=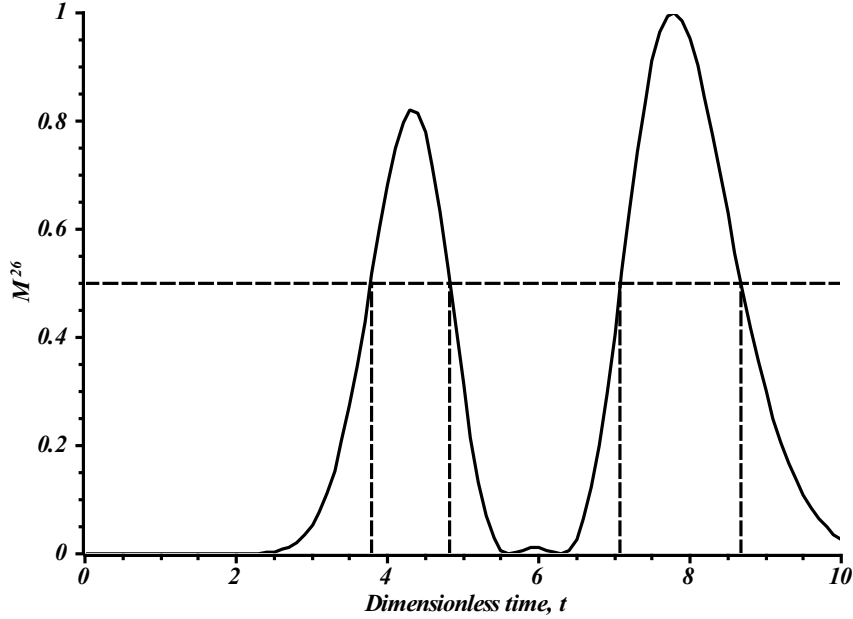,
  scale=0.6
   ,angle=270
}
\caption{The function $M^{26}(t)$ given by  eq.(\ref{hM}) with  
          $M^{26}_{max}=1.247\times 10^{-11}$ and $\displaystyle \varepsilon = \frac{\pi}{160}$;
          there are two time intervals 
          corresponding to $\displaystyle M^{26}(t)>\frac{1}{2}$: $ 3.781<t< 4.827$ and 
          $7.082<t<8.678$.
} 
  \label{Fig:E1min} 
\end{figure*}

Thus, for the  pair $(\varphi_2,\;\varphi_6)$, we have to determine such  time intervals and  region $\hat g$ in 
the space of parameters $\hat \varphi^A$  that 
both parameters with values from $\hat g$  may be detected at any instant during {{}these} time intervals. 
For the sake of simplicity, we  put other parameters $\varphi_i$ to zero.   We also take
\begin{eqnarray}\label{evex2}
\displaystyle\lambda^A_1=\lambda^A_2=\lambda^A_3=\frac{5}{16},
\end{eqnarray}
so that $\displaystyle \lambda^A_4=\frac{1}{16}$. In this case, only the second and the sixth columns  
of the matrix $H(t)$ (\ref{HH}) are nonzero, {{}and} we  may replace the matrix  
$H$  with the following   {{} $3\times 2$} matrix $\tilde H$:
\begin{eqnarray}
\tilde H(\varphi_2,\varphi_6, t)=\left(
\begin{array}{cc}
\displaystyle 
\frac{\partial a_0(\varphi_2,\varphi_6,t)}{\partial\varphi_2} &\displaystyle \frac{\partial a_0(\varphi_2,\varphi_6,t)}{\partial\varphi_6} \cr
\displaystyle \frac{\partial a_1(\varphi_2,\varphi_6,t)}{\partial\varphi_2} & 
\displaystyle \frac{\partial a_1(\varphi_2,\varphi_6,t)}{\partial\varphi_6} \cr
\displaystyle \frac{\partial a_2 (\varphi_2,\varphi_6,t)
}{\partial\varphi_2}&
\displaystyle \frac{\partial a_2(\varphi_2,\varphi_6,t) 
}{\partial\varphi_6}
\end{array}
\right).
\end{eqnarray}
Let us  consider the informational correlation over the time interval $0\le t \le T=10$ and define $\hat g(\varphi_2,\varphi_6) \subset \hat G^A$ as follows:
\begin{eqnarray}\label{gex1}
\hat g: \varepsilon <\varphi_2,\varphi_6<\frac{\pi}{2}-\varepsilon.
\end{eqnarray}
To find out the time intervals where both parameters $\varphi_i$ ($i=2,6$) may be detected, we 
construct the function ${{M}}^{26}(t)$ defined as 
\begin{eqnarray}\label{hM}
{{M}}^{26}(t)=\frac{M^{26}_2(t)}{M^{26}_{max}},
\;\;
M^{26}_2(t)=\min\limits_{\varphi_2,\varphi_6\in \hat g} \sum_{i=1}^3 |M^{26}_{2i}(\varphi_2,\varphi_6,t)|,\;\; M^{26}_{max} = \max\limits_{0\le t \le T} M^{26}_2(t),
\end{eqnarray}
where the minimization is over the parameters $\varphi_2$ and 
$\varphi_6$ inside of the region $\hat g$; $M^{26}_{2i}(\varphi_2,\varphi_6,t)$ ($i=1,2,3$) 
are the   second order minors of  the matrix $\tilde H$:
\begin{eqnarray}
\hspace{-1cm}
&&M^{26}_{21}=
\left|
\begin{array}{cc}
\displaystyle \frac{\partial a_0}{\partial\varphi_2} &
\displaystyle \frac{\partial a_0}{\partial\varphi_6} \cr
\displaystyle \frac{\partial a_1}{\partial\varphi_2}
&
\displaystyle \frac{\partial a_1}{\partial\varphi_6}
\end{array}
\right|,\;\;
M^{26}_{22}=
\left|
\begin{array}{cc}
\displaystyle \frac{\partial a_0}{\partial\varphi_2} &
\displaystyle \frac{\partial a_0}{\partial\varphi_6} \cr
\displaystyle \frac{\partial a_2}{\partial\varphi_2}&
\displaystyle \frac{\partial a_2}{\partial\varphi_6}
\end{array}
\right|,\;\;
M^{26}_{23}=
\left|
\begin{array}{cc}
\displaystyle \frac{\partial a_1}{\partial\varphi_2} &
\displaystyle \frac{\partial a_1}{\partial\varphi_6}\cr
\displaystyle \frac{\partial a_2}{\partial\varphi_2}
 &\displaystyle \frac{\partial a_2}{\partial\varphi_6}
\end{array}
\right|
,
\end{eqnarray}
where we do not write the parameters $\varphi_i$ in the arguments of the functions for the sake of simplicity.
The function  $ {{M}}^{26}(t)$ with $\displaystyle \varepsilon = \frac{\pi}{160}$   is depicted in Fig.\ref{Fig:E1min}. 
We see that the function  $ {{M}}^{26}(t)$ is positive during the rather long time 
intervals. However, the time intervals corresponding to  very small values 
of this function must be disregarded {{} because of the possible}  obstacles  
to detect parameters $\varphi_i$ ($i=2,6$) during these intervals (for instance , 
because of    fluctuations). 
For this reason, we suggest to use time intervals corresponding to $\displaystyle M^{26}>\frac{1}{2}$. Fig.\ref{Fig:E1min} shows that there are two such subintervals inside of the selected interval $0<t<10$: $ 3.781<t< 4.827$ and 
          $7.082<t<8.678$.   {{} Thus, we  calculate }
          the non-reducible informational correlation provided by the parameters $\varphi_2$ and $\varphi_6$ from $\hat g $ (\ref{gex1}) with 
zero values of  other parameters and eigenvalues  (\ref{evex2}): $E^{AB;min}(\hat g,t)=2$ inside of the  two 
above time subintervals 
(remember that $\displaystyle \lambda^B_i=\frac{1}{4}$, $i=1,2,3,4$, in this example because
${\mbox{ran}}\,
\hat T =7$). 

\paragraph{ Example 2:  ${\mbox{ran}} \,\hat T=9$,  $E^{AB}=6$ in Table 2.}
In {{} accordance with Table 2, $E^{AA}=6$.}  Using  formula (\ref{HH}) for $H$, we obtain that    the third order minors of the $3\times 6$ matrix $H$      are nonzero, so that  eq.(\ref{JH}) yields  $E^{AB;min}= 3$. The analysis of the third-order minors shows that any  triad of the parameters $\varphi_i$, $i=1,\dots,6$ may be transfered from the subsystem $A$ to the subsystem $B$ 
by the eigenvalues of the density matrix $\rho^B(\varphi^A,t)$ except for the triad of the parameters 
$\varphi_1,\varphi_3,\varphi_5$. {{} These parameters}
are introduced 
by the diagonal matrix  exponents $e^{i \gamma_3 \varphi_i}$ ($i=1,3,5$) in the unitary transformation (\ref{UrhoU3}) .

 Similar to the previous example,
 we  consider the informational correlation 
 established by {{} the one triad of the parameters, namely
 $\varphi_2$, $\varphi_4$, $\varphi_6$}, putting other parameters to zero.
 Thus the {{} restricted} set of  parameters is $\hat \varphi^A=\{\varphi_2, \varphi_4,\varphi_6\}$. 
 Therewith we take $\hat \varphi^A\in \hat G^A$ where 
 $\displaystyle\hat G^A: 0<\varphi_2,\varphi_4,\varphi_6<\frac{\pi}{2}$. 
We also fix eigenvalues as follows: 

\begin{eqnarray}\label{evex22}
&&
\displaystyle\lambda^A_1=\lambda^A_2=\lambda^A_3=\frac{4}{15}, \;\;
\displaystyle\lambda^A_4=\frac{1}{5},\\\nonumber
&&
\displaystyle\lambda^B_1=\lambda^B_2=\frac{4}{15}, \;\;\displaystyle\lambda^B_3=\displaystyle\lambda^B_4=\frac{7}{30}.
\end{eqnarray}

\begin{figure*}
   \epsfig{file=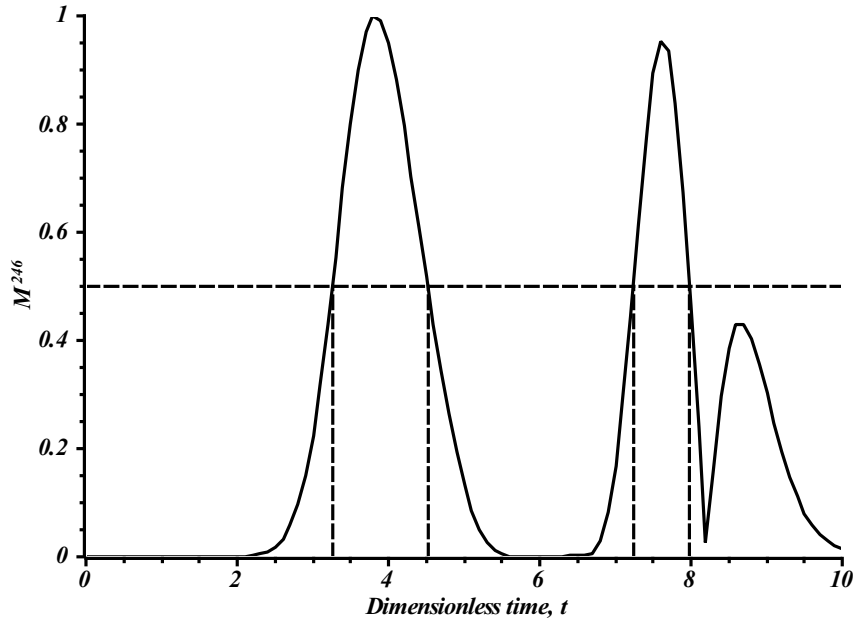,
  scale=0.6
   ,angle=270
}
\caption{The function $M^{246}(t)$ given by  eq.(\ref{hM2})) with
          $M^{246}_{max}=2.292\times  10^{-25}$, $\displaystyle \varepsilon = \frac{\pi}{50}$; there are two time intervals 
          corresponding to $\displaystyle M^{246}(t)>\frac{1}{2}$: $ 3.257<t< 4.520$ and 
          $7.233<t<7.983$}
  \label{Fig:E1min2} 
\end{figure*}

 In this case, only the second, the fourth and  the sixth columns of the matrix $H(t)$ (\ref{HH}) are nonzero 
 so that we  may replace $H$ by the following   {{} $3\times 3$} matrix $\tilde H$:
\begin{eqnarray}
\tilde H(\varphi_2,\varphi_4,\varphi_6,t)=\left(
\begin{array}{ccc}
\displaystyle 
\frac{\partial a_0(\varphi_2,\varphi_4,\varphi_6,t)}{\partial\varphi_2} &
\displaystyle \frac{\partial a_0(\varphi_2,\varphi_4,\varphi_6,t)}{\partial\varphi_4} &
\displaystyle \frac{\partial a_0(\varphi_2,\varphi_4,\varphi_6,t)}{\partial\varphi_6} \cr
\displaystyle \frac{\partial a_1(\varphi_2,\varphi_4,\varphi_6,t)}{\partial\varphi_2} & 
\displaystyle \frac{\partial a_1(\varphi_2,\varphi_4,\varphi_6,t)}{\partial\varphi_4} & 
\displaystyle \frac{\partial a_1(\varphi_2,\varphi_4,\varphi_6,t)}{\partial\varphi_6} \cr
\displaystyle \frac{\partial a_2 (\varphi_2,\varphi_4,\varphi_6,t)}{\partial\varphi_2}&
\displaystyle \frac{\partial a_2(\varphi_2,\varphi_4,\varphi_6,t) }{\partial\varphi_4}&
\displaystyle \frac{\partial a_2(\varphi_2,\varphi_4,\varphi_6,t) 
}{\partial\varphi_6}
\end{array}
\right).
\end{eqnarray}
Similar to the previous example, 
{{} we define the region $\hat g\subset\hat G^A$ as
\begin{eqnarray}\label{gex2}
 \hat g: \varepsilon < \varphi_2,\varphi_4,\varphi_6<\frac{\pi}{2} -\varepsilon.
 \end{eqnarray}
Let us consider the informational correlation 
during  the time interval $0\le t\le 10$ using  the parameters  
 inside of the region (\ref{gex2}). }
 To find out the time intervals where all three  parameters $\varphi_i$, $i=2,4,6$, 
 may be detected, we 
construct the function ${{ M}}^{246}(t)$ defined by the following formula  
(similar to eq.(\ref{hM})):
\begin{eqnarray}\label{hM2}
{{ M}}^{246}(t)=\frac{M^{246}_3(t)}{M^{246}_{max}},
\;\;
M^{246}_3(t)=\min_{\varphi_2,\varphi_4,\varphi_6\in \hat g} |M^{246}_{31}(\varphi_2,\varphi_4,\varphi_6,t)|,
\;\;
M^{246}_{max} = \max\limits_{0\le t \le T} M^{246}_3(t),
\end{eqnarray}
where the  minimization is over the parameters $\varphi_2$, $\varphi_4$ and $\varphi_6$ inside of the region $\hat g$, and 
the only nonzero 3rd order minor of $\tilde H$ is $\det\, \tilde H$, i.e.,
$M^{246}_{31}=\det\,\tilde H$. The function  ${{ M}}^{246}(t)$  with $\displaystyle \varepsilon =\frac{\pi}{50}$ 
is depicted in Fig.\ref{Fig:E1min2}. Similar to the previous example, we select the time intervals 
with $\displaystyle M^{246}>\frac{1}{2}$ as the suitable intervals  for the parameter detection. 
We see that there are two such subintervals inside of the interval $0<t<10$: $ 3.257<t< 4.520$ and  $7.233<t<7.983$. 
 {{} Thus we calculate } the non-reducible informational correlation provided by the parameters $\varphi_2$, $\varphi_4$ 
and $\varphi_6$ {{} from $\hat g$ (\ref{gex2})} with 
zero values of  other parameters and eigenvalues  (\ref{evex22}): $E^{AB;min}(\hat g,t)=3$ 
inside of the  two above time subintervals (remember that $\lambda^B_i$ satisfy either 
conditions (\ref{lam1},\ref{lam12}) or conditions (\ref{lam2},\ref{extra})). 

%%%%%%%%%%%%%%%%%%%%
\section{Informational correlations in long spin-1/2 chains:  particular examples}
\label{Section:long}

Results of Sec.\ref{Section:examples} allow us to conclude that 
 there are two basic obstacles in calculation of the informational correlation. 
First,
we have to calculate the Jacobian (\ref{JA}) using the formula (\ref{Jf}). The matrix dimensionality of 
this Jacobian increases with the increase in $N^B$ as $((N^A)^2 -1) \tilde D =((N^A)^2 -1)
N^A(N^A-1)$. 
Second, 
the calculation of the operator $T(t)$ in eq.(\ref{Jf})
is a very complicated dynamical problem for the  long chains. In fact, the dynamics of the  $N$-node
chain is described in the $2^N$-dimensional Hilbert space and 
may be simulated  only for small $N$ in general. However, 
the evolution of single excited spin along the spin-1/2 chain governed by the Hamiltonian commuting 
with the $z$-projection $I_z$ of the total spin momentum  is an exception.
In this case, the $N$-node chain evolves in the ($N+1$)-dimensional  Hilbert space. 

This chain 
was considered in \cite{Bose}, where, in addition, the approximation of nearest neighbor
interaction was used.   
Namely this model serves us as a model of 
the long spin-1/2 chain establishing the maximal possible informational correlation between the 
first and the last nodes which
we call subsystems $A$ and $B$, respectively 
(the remaining $N-2$ nodes compose the subsystem $C$ in this case).
Since $A$ is  a  one-node  subsystems, the  
group $SU(2)$ represents a group of unitary operators locally acting on this subsystem. 

Emphasize that the case of pure initial states considered in this section
allows us to simplify
formulas of Sec.\ref{Section:def} replacing the density matrix elements with the transition amplitudes. 
It is also important that, in the case of pure initial state with a
single  excitation in the subsystem $A$, the reduced  density matrix of the subsystem $A$ has 
only one non-zero eigenvalue and it equals to unit (i.e., the state of the subsystem $A$ is pure):
 $\Lambda={\mbox{diag}}(1,0,\dots,0)$. Consequently, such initial state of the subsystem $A$ 
 may be represented as 
 $\rho^A=U(\varphi^A)\Lambda U^+(\varphi^A)$. 
 Thus, all independent arbitrary parameters appearing below in the pure states 
 (\ref{initialB}) and (\ref{initialBB})
 are associated with  parameters of the unitary transformation.

Let us use the following conventional basis  \cite{Bose}:
\begin{eqnarray}\label{basisB}
|\mathbf{j}\rangle,\;\;\mathbf{j} =\mathbf{0},\mathbf{1},\dots,\mathbf{N},
\end{eqnarray}
where ${j}$ means the location of the excited node 
(i.e., the $j$th node is  directed opposite to the external magnetic field) while all other nodes 
are in the basic state (i.e., they are directed along the external field). 
Therewith $\mathbf{0}$ means that all nodes are in the basic state. 
If we deal with a single node (subsystems $A$ or $B$)
 we use notation $|0\rangle$ for the spin in the basic state and $|1\rangle$ 
 for the excited spin.

Two transferable parameters of the $SU(2)$ are incorporated into the arbitrary initial state of the 
 first spin \cite{Bose} as follows:
\begin{eqnarray}\label{initialB}
|\psi_1(0)\rangle = \cos\frac{\varphi_1}{2} | 0\rangle  + 
e^{i\varphi_2} \sin\frac{\varphi_1}{2} | 1\rangle.
\end{eqnarray}
The initial density matrix $\rho^A(\varphi^A)$ reads
\begin{eqnarray}
\rho^A(\varphi^A) = |\psi_1(0)\rangle \langle \psi_1(0)|.
\end{eqnarray}
Thus, $E^{AA}=2$, $E^{AA}_{norm}=1$, i.e., the information encoded into the subsystem $A$ reaches  
its  maximal possible value as is shown in Sec.\ref{Section:def}, eq.(\ref{DA})
(see also Sec.\ref{Section:EAA1n}). Although the initial state of the whole system is pure, the state of 
the $N$th node is a mixed one and is described by the  following
  reduced density matrix \cite{Bose}:
\begin{eqnarray}\label{lnode}
\rho^B(t)=P(t)|\psi^B(t)\rangle \langle \psi^B(t)| +
(1-P(t))  |0\rangle \langle 0|,
\end{eqnarray}
where 
\begin{eqnarray}
&&|\psi^B(t)\rangle =  
\frac{1}{\sqrt{P(t)}} \left(a_1(\varphi_1) | 0\rangle  + a_2(\varphi_1,\varphi_2,t) | 1\rangle\right),
\\\nonumber
&&
a_1(\varphi_1) = \cos\frac{\varphi_1}{2}, \;\;a_2(\varphi_1,\varphi_2,t)=
e^{i\phi_2} \sin\frac{\varphi_1}{2} f^N_{AB}(t),\\\nonumber
&&
P(t)=\cos^2\frac{\varphi_1}{2} + \sin^2\frac{\varphi_1}{2} |f^N_{AB}(t)|^2,\\\nonumber
&&
f^N_{AB}(t)= \langle \mathbf{N}| e^{-i H t} |\mathbf{1}\rangle.
\end{eqnarray}
Here $f^N_{AB}(t)$ is the transition amplitude of an excitation from the 1st to the $N$th 
spin, $|\mathbf{1}\rangle$ and $|\mathbf{N}\rangle$ mean the states with the first 
and last excited spin respectively (in accordance with eq.(\ref{basisB})), 
and $H$ is the Hamiltonian governing the dynamics of 
the spin chain with the only restriction $[H,I_z]=0$. 
Having  the parameters $a_i$, $i=1,2$,
we may find two  real parameters $\varphi_i$, $i=1,2$ solving the following system 
of one real and one complex equations (i.e., the system is overdetermined but compatible):
\begin{eqnarray}\label{aa}
a_1=\cos\frac{\varphi_1}{2}, \;\; 
a_2= e^{i\phi_2} \sin\frac{\varphi_1}{2} f^N_{AB}(t).
\end{eqnarray}
This system is
solvable at any instant satisfying the condition
\begin{eqnarray}
f^N_{AB}(t) \neq 0.
\end{eqnarray} 
Therewith $P(t)>0$ by definition.
Thus, the maximal possible informational correlation $E^{AB}=2$ ($E^{AB}_{norm}=1$) 
is established between one-node 
subsystems $A$ and $B$ (the first and the last nodes of the $N$-node spin chain).
At such time moments  that $f^N_{AB}(t) = 0$, only the first of eqs.(\ref{aa})
remains, so that $E^{AB}=1$ ($E^{AB}_{norm}=1/2$).

Now we show how the system with a single excited node may be 
used to establish the informational correlation between
the large subsystems $A$ and $B$. 
It is convenient to use the following notations for the basis vectors of the $M$-node subsystem.
\begin{eqnarray}\label{basisB2}
|\mathbf{j}_M\rangle,\;\;\mathbf{j} =\mathbf{0},\mathbf{1},\dots,\mathbf{M},
\end{eqnarray}
where we explicitly use the length $M$ of the selected subsystem.
Let the subsystems $A$ and $B$ of the $N$-node spin chain consist of $\tilde N^A$ first nodes 
and $\tilde N^B$ last nodes respectively.
We generalize the initial state (\ref{initialB}) as follows:
\begin{eqnarray}\label{initialBB}
&&|\psi^A(0)\rangle = \alpha_0 |\mathbf{0}_{\tilde N^A}\rangle  + \sum_{j=1}^{\tilde N^A} 
\alpha_j|\mathbf{j}_{\tilde N^A} \rangle,
\end{eqnarray}
where $\alpha_0$ may be taken as a real parameter and  parameters $\alpha_j$, $j>0$, are the
complex ones with the normalization
\begin{eqnarray}\label{normAN}
\alpha_0^2 +\sum_{i=1}^{\tilde N^A} |\alpha_i|^2 =1.
\end{eqnarray}
We require that all other spins are in the basic state, i.e., the initial state of the whole system reads:
\begin{eqnarray}\label{inst}
|\psi(0)\rangle = |\psi^A(0)\rangle \otimes |\mathbf{0}_{N-\tilde N^A}\rangle,
\end{eqnarray}
where
 $|\mathbf{0}_{N-\tilde N^A} \rangle$ means the basic state of the 
 chain of ${N-\tilde N^A}$ nodes.

Thus, all in all, we have $2\tilde N^A$ arbitrary real parameters: $2\tilde N^A+1 $ 
parameters $a_0$,  ${\mbox{Re}}\alpha_i$, 
$ {\mbox{Im}}\alpha_i$, $i=1,\dots,\tilde N^A$ related by the normalization condition (\ref{normAN}). 
As shown in the beginning of this section, 
these arbitrary parameters are associated with  $2\tilde N^A$ arbitrary parameters $\varphi_i$ 
of the unitary transformation. For simplicity, 
we do not explicitly write this dependence in the arguments of $\alpha_i$. 
Thus,  $E^{AA}=2 \tilde N^A$, 
$\displaystyle E^{AA}_{norm}=\frac{2 \tilde N^A}{2^{\tilde N^A} (2^{\tilde N^A}-1)}$.
Of course, the found $E^{AA}$ is less then the 
maximal possible  number of parameters that may be encoded  into the mixed initial 
state of the subsystem $A$ consisting of $\tilde N^A$ nodes
considered in Sec.\ref{Section:def}. Thus, the subsystems $A$ and $B$ may 
not possess the maximal possible informational correlation (which is 
$E^{AA}=2^{\tilde N^A}(2^{\tilde N^A}-1)$) using the pure initial state (\ref{initialBB})
and the Hamiltonian commuting with $I_z$. 

With time, the initial state of the whole system  (\ref{inst}) evolves as follows:
\begin{eqnarray}
|\psi(t)\rangle = e^{-iH t} |\psi^A(0)\rangle \otimes |\mathbf{0}_{N-\tilde N^A} \rangle.
\end{eqnarray}

 The state of the subsystem $B$ is described by the following reduced density matrix 
 \begin{eqnarray}\label{lnode2}
\rho^B_{\tilde N^B}(t)=P_{\tilde N^B}(t)|\psi^B_{\tilde N^B}(t)\rangle \langle \psi^B_{\tilde N^B}(t)| +
(1-P_{\tilde N^B}(t))  |0\rangle \langle 0|
\end{eqnarray}
where 
\begin{eqnarray}
|\psi^B_{\tilde N^B}(t) \rangle = 
 \frac{1}{\sqrt{ P_{\tilde N^B}(t)}}
 \left(\beta_0 |\mathbf{0}_{\tilde N^B} \rangle +
\sum_{j=1}^{N^B} \beta_j |\mathbf{j}_{\tilde N^B} \rangle \right),
\end{eqnarray}
 $\beta_i$ are the  transition amplitudes:
\begin{eqnarray}\label{bet0}
\beta_0 &=& 
\left( \langle \mathbf{0}_{N-\tilde N^B}| \otimes \langle \mathbf{0}_{\tilde N^B}| \right)|
 \psi(t)\rangle =\alpha_0\\\label{betj}
 \beta_j &=& 
 \left(\langle \mathbf{0}_{N-\tilde N^B}| \otimes \langle \mathbf{j}_{\tilde N^B}|\right)|\psi(t)\rangle =
\sum_{k=1}^{\tilde N^A} \alpha_k r_{jk} ,\\\label{rjk}
&&
r_{jk}=  \langle \mathbf{j}_{\tilde N^B}| \otimes\langle \mathbf{0}_{N-\tilde N^B} |
e^{-iHt }|\mathbf{k}_{\tilde N^A}\rangle \otimes|\mathbf{0}_{N-\tilde N^A}\rangle,
\end{eqnarray}
and $ P_{\tilde N^B}(t) $ is the normalization:
\begin{eqnarray}
 \label{normBN}
&&
 P_{\tilde N^B}(t) = \beta_0^2 +\sum_{j=1}^{N^B} |\beta_j|^2. 
\end{eqnarray}
Thus, we may register at most $2\tilde N^B$ real parameters in the subsystem $B$, i.e., 
$E^{AB}(t)\le 2\tilde N^B$,$\displaystyle E^{AB}_{norm}(t)\le \frac{2\tilde N^B}{2^{\tilde N^B} (2^{\tilde N^B}-1)}$.
Let us use the transition amplitudes $\beta_j$, $j=0,1,\dots,\tilde N^B$, rather then the 
density matrix elements to calculate  the number of transfered  parameters $\varphi_i$
 and write system (\ref{bet0},\ref{betj}) in the matrix form:
 \begin{eqnarray}\label{bTa}
\boldsymbol{\beta} = \boldsymbol{ T\,\alpha},
 \end{eqnarray}
 where
 \begin{eqnarray}
 &&\boldsymbol{\beta} =(\beta_0,\beta_1,\dots,\beta_{\tilde N^B})^T,\;\;
\boldsymbol{ \alpha} =(\alpha_0,\dots,\alpha_{\tilde N^A})^T,\;\;\\\nonumber
&& 
 \mathbf{T}= \left(
\begin{array}{cccc}
1&0&\cdots&0\cr
0&r_{11}&\cdots & r_{1\tilde N^A}\cr
\cdots &\cdots &\cdots &\cdots \cr
0&r_{\tilde N^B 1}&\cdots & r_{\tilde N^B\tilde N^A}
\end{array}  
 \right).
 \end{eqnarray}
 Equation (\ref{bTa}) is the  analogue of equation (\ref{TTmel2}) in  Sec.\ref{Section:def}.
Let $\tilde N^A=\tilde N^B$ hereafter in this section. Then, $\mathbf{T}$ is a square matrix and  the condition
 \begin{eqnarray}\label{detlong}
 \det \boldsymbol{T} \neq 0
 \end{eqnarray}
 provides the complete information transfer. Consequently,  the 
 maximal informational correlation between the subsystems $A$ and $B$ is reached: 
 $E^{AB}_{max} = 2 \tilde N^A$, 
 $\displaystyle E^{AB}_{norm}(t)= \frac{2\tilde N^A}{2^{\tilde N^A} (2^{\tilde N^A}-1)}$. 
 
 For the further study, let us split the real and imaginary parts of 
 eq. (\ref{bTa}) and rewrite this equation as 
\begin{eqnarray}\label{alpbet}
\left(\begin{array}{c}
{\mbox{Re}}\; \boldsymbol{\beta}\cr
{\mbox{Im}}\; \boldsymbol{\beta}
\end{array}\right)= \hat {\boldsymbol{T}}
 \left(\begin{array}{c}
{\mbox{Re}}\; \boldsymbol{\alpha}\cr
{\mbox{Im}}\; \boldsymbol{\alpha}
\end{array}\right),\;\;\; \hat {\boldsymbol{T}} =\left(
\begin{array}{cc}
{\mbox{Re}}\; \boldsymbol{T} & - {\mbox{Im}}\; \boldsymbol{T} \cr
{\mbox{Im}}\; \boldsymbol{T} & {\mbox{Re}}\; \boldsymbol{T}
\end{array}
\right).
\end{eqnarray}
Eq.(\ref{alpbet}) is an analogue of eq.(\ref{RITf}) in Sec.\ref{Section:def}.
 We see, that the number of real equations in system (\ref{alpbet}) is  $2 \tilde N^A +1$. This
 is one grater then the 
 number of 
 independent parameters $\varphi_i$
 that may be found from system  (\ref{alpbet}).  
 Thus, similar to the mixed initial state case considered in Sec.\ref{Section:def}, 
 condition (\ref{detlong}) is enough, but it is not a necessary  condition for the maximal possible
 informational correlation.
 If $\det \boldsymbol{T} =0$,
 then the parameter $E^{AB}$ is defined by the rank of $\boldsymbol{T}$. In turn, this rank 
 is defined by the particular choice of the spin chain length and Hamiltonian.
 Formally, the maximal possible informational correlation
 $E^{AB}$ may be achieved for ${\mbox{ran}}\,\hat {\boldsymbol{T}} \ge 2 \tilde N^A$. Otherwise, we 
 deal with the partial information correlation. Further 
 examples of informational correlations in quantum systems will be given in different paper.

 {\bf Non-reducible informational correlation.}
 Regarding the non-reducible informational correlation, it is unit in the considered examples. In fact, both 
 reduced density matrices  (\ref{lnode}) and (\ref{lnode2}) are written in the diagonal form
  and have two nonzero eigenvalues:
 $P, (1-P)$ and $P_{\tilde N^B}, (1-P_{\tilde N^B})$ respectively.  
 Thus only one parameter may be transfered by the eigenvalues of the reduced 
 density matrix associated with the subsystem $B$, i.e.,
 $E^{AB;min} =1$, $\displaystyle E^{AB;min}_{norm} =\frac{1}{2^{\tilde N^A} (2^{\tilde N^A}-1)}$.
 
%%%%%%%%%%%%%%%%%
\section{Conclusions}
\label{Section:conclusions}
We  introduce the informational correlation between two subsystems {{} $A$ and $B$} 
as the possibility to effect on the state of the subsystem $B$ through the parameters of the unitary 
transformation $U^A$  locally performed on the subsystem $A$ and vice-versa. The measure of the 
informational correlation $E^{AB}$  {{} equals to} the number of  parameters of the local unitary transformation $U^A$ 
which may be detected in the subsystem $B$. 
We also introduce the normalized measure of the informational correlation $E^{AB}_{norm}$ showing whether 
the informational correlation is far from the saturation. The so-called non-reducible informational correlation 
$E^{AB;min}(t)$ is of a  special interest, because this part of informational correlation is invariant with 
respect to the local unitary transformations of the subsystem $B$ at the time  instant $t$. 
 
 {{} 
Below we represent the list of the basic properties of 
 the informational correlation (for the tensor product  initial 
density matrix (\ref{inden})) and compare them with the analogous properties of the discord and 
 entanglement (if this is possible).}

\begin{enumerate}
\item
Unlike the entanglement and discord, the informational correlation represents a dynamical characteristics which is identical to zero at the initial time instant. 
\item
 {{} 
By its definition (\ref{EAB}) in terms of the rank of some Jacobian matrix, 
the informational correlation takes the discrete set of values, 
unlike the entanglement and discord. Moreover, this definition provides 
the stability of the informational correlation 
with respect to deviations of the system's parameters (such as the dipole-dipole interaction 
constants and the magnetic field distribution).}
\item
The informational correlation is invariant with respect to the initial local unitary transformations of the subsystem $A$, similar to the usual entanglement and discord. 
However, the informational correlation is not invariant with respect to the local unitary transformations of the subsystem $B$ (either initial or $t$-dependent), unlike the entanglement and discord. Consequently,
using the local unitary transformations of the receiver $B$  we may handle (up to a certain extent) the number of the parameters  transfered from the subsystem $A$ to the subsystem $B$ and, thus, manipulate the informational correlation $E^{AB}$. The local transformations performed on the subsystem $C$ may also effect $E^{AB}$.
\item
$E^{AB}(t) \equiv 0$ only if the  initial density matrix $\rho^A(0)$ in formula (\ref{inden})  is proportional to the identity matrix.   For the tensor product initial state (\ref{inden},\ref{inden0}), $ E^{AB}=E^{BA} =0$ only if both $\rho^A(0)$ and $\rho^B(0)$ are proportional to the identity matrix. The unitary invariant discord possesses the same property \cite{Z_Disc}.
\item
The complete information transfer is not required in order to obtain the maximal possible value of $E^{AB}$, because the maximal possible number of arbitrary  parameters $\varphi_i$ transfered  from  $A$ to $B$ is less then $(N^A)^2-1$ (the maximal number of different real parameters in the $N^A\times N^A$ density  matrix).
\item
The informational correlation is  sensitive to the  multiplicity of the  eigenvalues of the matrices $\rho^A(0)$ and $\rho^B(0)$ for the case of the tensor product initial state (\ref{inden},\ref{inden0}).

\item
It is interesting that the conditions $E^{AA} < \tilde D^A$ and $E^{AB} < E^{AA}$
require the  strong relations among  the eigenvalues $\lambda^A_i$, $\lambda^B_i$ and $\lambda^C_i$. 
For  the particular examples, these relations  have been found in  
Sec.\ref{Section:examples}, see eqs.(\ref{ran},\ref{lamB1},\ref{lam1},\ref{lam2}) and Tables 1,2. 
The minor deviation from these exact relations  leads  (i) to the encoding of the maximal possible parameters 
$\tilde D^A$ into the subsystem $A$ and (ii) to the spread of the complete information throughout 
the whole system and consequently to the  maximal possible informational correlation
$E^{AB}=E^{AA}=\tilde D^A$. {{} This phenomenon was not observed in the case of entanglement and discord.
Presumably, such behavior of a system 
 must be  closely related with the  fluctuations of the informational correlation 
 and requires  the
more detailed  study. }
\item
There are two subsets of parameters $\varphi_i$ transfered from $A$ to $B$: $\varphi^U$ and $\varphi^\Lambda$.
The first one may be detected in the matrix of the eigenvectors of the reduced density matrix $\rho^B(\varphi^A,t)$, while the second subset is transfered by the eigenvalues of the same matrix. The  subset  $\varphi^\Lambda$
is most reliable for the purpose of the information transfer, because the number of parameters in this subset may not be decreased by any local unitary transformation performed on the subsystem $B$. Namely this subset  is responsible for the non-reducible informational correlation $E^{AB;min}$. Note that some of the parameters $\varphi_i$ might be encoded in both subsets $\varphi^U$ and $\varphi^\Lambda$.  The informational correlation $E^{AB}$ and the non-reducible 
informational correlation $E^{AB;min}$ 
might be viewed as the analogues of the total and the classical 
correlations in the definition of the discord. The removable informational correlation 
$\Delta E^{AB}=E^{AB}-E^{AB;min}$  is the analogue of the discord itself.
\item
{{}
Presumably,
the case of non-separable initial state will reveal some additional 
interesting features. For instance, the 
informational correlations established by the subgroup of LNUs \cite{F,GKB,DG} should 
be studied. }

\item 
{{}
Parameters of the group $SU(N^A)$ transfered form $A$ to $B$ may 
be treated as bits of   information. Local transformations of the subsystems $B$ 
(and perhaps $C$)
provide a control of information transfer. These two facts suggest us to consider a bi-partite
(or three-partite) quantum system with local transformations 
as a controllable  gate or chain of gates (if the subsystems are large enough). 
In the case of 1-spin
subsystems $A$ and $B$ we have 2 parameters (which might be treated, for instance, as 2 bits). The 2-spin subsystems provide us with 12 bits. 
In general, $n$-node subsystems $A$ and $B$ of spin-1/2 particles 
yield $(2^{2n} - 2^n)$ bits. The introduced measure $E^{AB}$ allows us to handle the remote control 
of gates. Both the  gate construction and the remote control of gates
are  those problems that deserve the further study. 
}
\item
We represent the detailed study of the informational correlations 
(including the non-reducible ones)  between the one- and 
two-node subsystems $A$ and $B$ 
in the four node spin-1/2 chain  with mixed initial states governed by the XY Hamiltonian,
Sec.\ref{Section:examples}. 
The informational correlation may be relatively simply calculated 
in the spin-1/2 chains having a pure initial state with single initially excited node 
of the subsystem $A$ provided that the spin dynamics is governed  by the Hamiltonian 
commuting with $I_z$ ( $z$-projection of the total spin momentum), as is shown 
in Sec.\ref{Section:long}.
The case of one-particle subsystems $A$ and $B$ is  studied in details; general 
formulas for the calculation of informational correlation between arbitrary subsystems $A$ and $B$ 
of long  chain are derived.
The non-reducible informational correlation equals one for this type of initial states.
\end{enumerate}

{{}
Finally we would like to notice that, although there is no apparent relation between 
the informational correlation and LE, the idea of increasing  the amount of correlations between two selected 
particles by doing 
local measurements on the rest of quantum system  \cite{VPC,JK} may be used 
 in the further development of application 
 of informational 
 correlation.
}

Author thanks Profs. E.B.Fel'dman, M.A.Yurishchev and Dr. A.N.Pyrkov for useful discussion. 
Author also thanks reviewer for useful comment. 
This work is partially supported by the Program of the 
Presidium of RAS No.8 ''Development of methods of obtaining chemical 
compounds and creation of new materials'' and by the RFBR grant No.13-03-00017

\section{Appendix}
\label{Section:appendix}

\subsection{A. Explicit form of the matrices $\gamma_i$. }
\label{Section:A}
We give the list of matrices $\gamma_i$ representing the  basis of the Lie algebra of  $SU(4)$ \cite{GM}:
\begin{eqnarray}\label{gamma}
&&
\gamma_1=\left[
\begin{array}{cccc}
         0& 1& 0& 0\cr
         1& 0& 0& 0\cr
         0& 0& 0& 0\cr
         0& 0& 0& 0  
\end{array}
\right],\;\;
\gamma_2=\left[
\begin{array}{cccc}
         0&-i& 0& 0\cr
         i& 0& 0& 0\cr
         0& 0& 0& 0\cr
         0& 0& 0& 0  
\end{array}
\right],\;\;
\gamma_3=\left[
\begin{array}{cccc}
         1& 0& 0& 0\cr
         0&-1& 0& 0\cr
         0& 0& 0& 0\cr
         0& 0& 0& 0  
\end{array}
\right],\\\nonumber
&&
\gamma_4=\left[
\begin{array}{cccc}
         0& 0& 1& 0\cr
         0& 0& 0& 0\cr
         1& 0& 0& 0\cr
         0& 0& 0& 0  
\end{array}
\right],\;\;
\gamma_5=\left[
\begin{array}{cccc}
         0& 0&-i& 0\cr
         0& 0& 0& 0\cr
         i& 0& 0& 0\cr
         0& 0& 0& 0  
\end{array}
\right],\;\;
\gamma_6=\left[
\begin{array}{cccc}
         0& 0& 0& 0\cr
         0& 0& 1& 0\cr
         0& 1& 0& 0\cr
         0& \;\;0& 0& 0  
\end{array}
\right],\\\nonumber
&&
\gamma_7=\left[
\begin{array}{cccc}
         0& 0& 0& 0\cr
         0& 0&-i& 0\cr
         0& i& 0& 0\cr
         0& 0& 0& 0  
\end{array}
\right],\;\;
\gamma_8=\frac{1}{\sqrt{3}}\left[
\begin{array}{cccc}
         1& 0& 0& 0\cr
         0& 1& 0& 0\cr
         0& 0&-2& 0\cr
         0& 0& 0& 0  
\end{array}
\right],\;\;
\gamma_9=\left[
\begin{array}{cccc}
         0& 0& 0& 1\cr
         0& 0& 0& 0\cr
         0& 0& 0& 0\cr
         1& 0& 0& 0  
\end{array}
\right],\\\nonumber
&&
\gamma_{10}=\left[
\begin{array}{cccc}
         0& 0& 0&-i\cr
         0& 0& 0& 0\cr
         0& 0& 0& 0\cr
         i& 0& 0& 0  
\end{array}
\right],\;\;
\gamma_{11}=\left[
\begin{array}{cccc}
         0& 0& 0& 0\cr
         0& 0& 0& 1\cr
         0& 0& 0& 0\cr
         0& 1& 0& 0  
\end{array}
\right],\;\;
\gamma_{12}=\left[
\begin{array}{cccc}
         0& 0& 0& 0\cr
         0& 0& 0&-i\cr
         0& 0& 0& 0\cr
         0& i& 0& 0  
\end{array}
\right],\\\nonumber
&&
\gamma_{13}=\left[
\begin{array}{cccc}
         0& 0& 0& 0\cr
         0& 0& 0& 0\cr
         0& 0& 0& 1\cr
         0& 0& 1& 0  
\end{array}
\right],\;\;
\gamma_{14}=\left[
\begin{array}{cccc}
         0& 0& 0& 0\cr
         0& 0& 0& 0\cr
         0& 0& 0&-i\cr
         0& 0& i& 0  
\end{array}
\right],\;\;
\gamma_{15}=\frac{1}{\sqrt{6}}\left[
\begin{array}{cccc}
         1& 0& 0& 0\cr
         0& 1& 0& 0\cr
         0& 0& 1& 0\cr
         0& 0& 0&-3  
\end{array}
\right].
\end{eqnarray}

\subsection{ B. Proof of eq.(\ref{JH}) for multiple and zero eigenvalues $\lambda_i$ of the matrix $\rho^B(\varphi^A,t)$}
\label{Section:B}

Suppose that there are $(P+1)<(N^B-1)$ nonzero different eigenvalues of the matrix $\rho^B(\varphi,t)$. Then, in order to 
define the non-reducible informational correlation $E^{AB;min}$, we 
may replace the characteristic equation (\ref{char}) with 
the following polynomial one:
\begin{eqnarray}\label{Bchar}
\prod_{i=1}^P (\lambda-\lambda_i(\varphi^A,t)) = \lambda^P +\sum_{j=0}^{P-1} \tilde a_i(\varphi^A,t) \lambda^i=0,
\end{eqnarray}
where $\tilde a_i$ are expressed in terms of $\lambda_i$.
In this equation, we take into account only $P$ different nonzero  eigenvalues because of the identity 
$\displaystyle {\mbox{Tr}}\rho^B=\sum\limits_{i=1}^{P+1} q_i \lambda_i=1$, where $q_i$ is the multiplicity of the root $\lambda_i$. 
Then the non-reducible informational correlation is defined by the rank of the matrix 
\begin{eqnarray}
 J^B_\Lambda(\varphi^A,t)=\frac{\partial (\lambda_1(\varphi^A,t),\dots, \lambda_P(\varphi^A,t))}
{\partial(\varphi_1,\dots,\varphi_{\tilde D^A})},
\end{eqnarray}
so that
\begin{eqnarray}\label{BnonredE}
E^{AB;min}(\varphi^A,t) = {\mbox{ran}}\, J^B_\Lambda(\varphi^A,t).
\end{eqnarray}
Differentiating eq.(\ref{Bchar})  with respect to the parameters $\varphi_k$, $k=1,\dots,\tilde D^A$, and 
solving the resulting equations for $\displaystyle \frac{\partial\lambda}{\partial \varphi_k}$ we obtain:
\begin{eqnarray}
\frac{\partial\lambda}{\partial \varphi_k} =
-\frac{\displaystyle \sum_{i=0}^{P-1} \frac{\partial \tilde a_i(\varphi^A,t)}{\partial\varphi_k} \lambda^i}
{\displaystyle P\lambda^{P-1} + \sum_{i=1}^{P-1}  i \tilde a_i(\varphi^A,t) \lambda^{i-1}}.
\end{eqnarray}
Therefore, for the matrix $ J^B_\Lambda(\varphi^A,t)$ one has
\begin{eqnarray}\nonumber
J^B_\Lambda(\varphi^A,t)&=&\frac{1}
{\tilde J_0(\varphi^A,t)}\left(
\begin{array}{ccc}\displaystyle
\sum_{i=0}^{P-1}  \frac{\partial \tilde  a_i(\varphi^A,t)}{\partial\varphi_1} \lambda^i_1(\varphi^A,t)&\cdots&\displaystyle
\sum_{i=0}^{P-1}  \frac{\partial\tilde a_i(\varphi^A,t)}{\partial\varphi_{\tilde D^A}} \lambda^i_{1}(\varphi^A,t)\cr
\cdots&\cdots&\cdots \cr\displaystyle
\sum_{i=0}^{P-1}  \frac{\partial \tilde a_i(\varphi^A,t)}{\partial\varphi_{1}} \lambda^i_P(\varphi^A,t)&\cdots&\displaystyle
\sum_{i=0}^{P-1}  \frac{\partial \tilde a_i(\varphi^A,t)}{\partial\varphi_{\tilde D^A}} \lambda^i_{P}(\varphi^A,t)
\end{array}
\right) =\\\label{BLamH}
&&\tilde \Lambda^B(\varphi^A,t) \tilde H(\varphi^A,t) ,
\end{eqnarray}
where 
\begin{eqnarray}
\tilde J_0(\varphi^A,t)=(-1)^P
\prod_{j=1}^P \Big(P\lambda^{P-1}_j(\varphi^A,t) + \sum_{i=1}^{P-1}  i \tilde a_i(\varphi^A,t) \lambda^{i-1}_j(\varphi^A,t)\Big) \neq 0,
\end{eqnarray}
while $\tilde \Lambda^B$ and $\tilde H$ are the $P\times P$ and  $P\times \tilde D^A$  matrices respectively:
\begin{eqnarray}\label{BLam}
&&\tilde\Lambda^B=\left(
\begin{array}{cccc}\displaystyle
  1 &\lambda_1(\varphi^A,t)& \cdots&
\lambda_1^{P-1}(\varphi^A,t)\cr
\cdots&\cdots&\cdots&\cdots\cr
  1 &\lambda_P(\varphi^A,t)& \cdots&
\lambda_P^{P-1}(\varphi^A,t)
\end{array}
\right),\\ \label{BHH}
&&\tilde H(\varphi^A,t)=\left(
\begin{array}{ccc}\displaystyle
  \frac{\partial \tilde a_0(\varphi^A,t)}{\partial\varphi_1} &\cdots&
  \displaystyle\frac{\partial \tilde a_0(\varphi^A,t)}{\partial\varphi_{\tilde D^A}}
  \cr
  \cdots&\cdots&\cdots \cr\displaystyle
\displaystyle
  \frac{\partial \tilde a_{P-1}(\varphi^A,t)}{\partial\varphi_1} 
  &\cdots&\displaystyle
  \frac{\partial \tilde a_{P-1}(\varphi^A,t)}{\partial\varphi_{\tilde D^A}} 
\end{array}
\right) .
\end{eqnarray}
Then eq.(\ref{BnonredE})  yields
\begin{eqnarray}\label{BJH}
E^{AB;min}(\varphi^A,t) ={\mbox{ran}} \, J^B_\Lambda(\varphi^A,t)={\mbox{ran}} \,\tilde  H(\varphi^A,t).
\end{eqnarray}
 Now notice that $P$ coefficients $\tilde a_i$ in eq.(\ref{Bchar}) are defined by $P$ different nonzero eigenvalues $\lambda_i$, $i=1,\dots,P$. 
From another hand, the
coefficients $a_i$ in eq.(\ref{char}) are defined by the same   $P$ independent eigenvalues $\lambda_i$, $i=1,\dots,P$, 
and consequently by $P$ coefficients $\tilde a_i$, $i=1,\dots,P$. 
The last statement is provided by the relation between sets 
$\tilde a_i$ and $\lambda_i$. This relation follows from eq.(\ref{Bchar}), 
{{} where 
all $\lambda_i$, $i=1,\dots,\lambda_P$,  are different by our requirement. Consequently,}
\begin{eqnarray}\label{tal}
\left|
\frac{\partial (\tilde a_0,\dots,\tilde a_{P-1})}{\partial(\lambda_1,\dots,\lambda_P)}
\right|\neq 0.
\end{eqnarray}
Thus, for the matrix $H$ represented by eq.(\ref{HH}),  we may write 
\begin{eqnarray}\label{HFH}
H(\varphi^A,t) = F(\varphi^A,t) \tilde H(\varphi^A,t),
\end{eqnarray}
where $F$ is $(N^B-1) \times P$ matrix,
\begin{eqnarray}\label{F}
F(\varphi^A,t)=\left(
\begin{array}{ccc}\displaystyle
  \frac{\partial  a_0(\varphi^A,t)}{\partial \tilde a_0} &\cdots&\displaystyle
  \frac{\partial  a_0(\varphi^A,t)}{\partial \tilde a_{P-1}} \cr
\cdots&\cdots&\cdots \cr\displaystyle
\frac{\partial  a_{N^B-2}(\varphi^A,t)}{\partial \tilde a_{0}} &\cdots&\displaystyle
  \frac{\partial  a_{N^B-2}(\varphi^A,t)}{\partial \tilde a_{P-1}}
\end{array}
\right) .
\end{eqnarray}
It may be readily shown that  the rank of the matrix $F$ takes its maximal possible value, ${\mbox{ran}}\, F = P$,  $P\le (N^B-1)$. 
In fact, since $a_i$, $i=0,1,\dots, N^B-2$, are expressed in terms of $\lambda_i$ (see eqs.(\ref{char0},\ref{char})) and there are only $P$ independent eigenvalues $\lambda_i$, $i=1,\dots,P$, then
${\mbox{ran}}\displaystyle \frac{\partial(a_0,a_1,\dots,a_{N^B-2})}{\partial(\lambda_1,\dots,\lambda_P)}=P$. But
$\displaystyle \frac{\partial(a_0,a_1,\dots,a_{N^B-2})}{\partial(\lambda_1,\dots,\lambda_P)} =
F \frac{\partial (\tilde a_0,\dots,\tilde a_{P-1})}{\partial(\lambda_1,\dots,\lambda_P)}$. Consequently, in virtue of condition (\ref{tal}), we conclude that $\displaystyle {\mbox{ran}}\, F ={\mbox{ran}}\displaystyle \frac{\partial(a_0,a_1,\dots,a_{N^B-2})}{\partial(\lambda_1,\dots,\lambda_P)}=P$.
Thus  the rank of the product $F \tilde H$ equals to the rank of $ \tilde H$ in eq.(\ref{HFH}), which yields
\begin{eqnarray}\label{JHA}
{\mbox{ran}}\, H(\varphi^A,t) = {\mbox{ran}}\, \tilde H(\varphi^A,t).
\end{eqnarray}
In turn, eq.(\ref{JHA}) means that eq.(\ref{JH}) holds for the multiple and/or zero eigenvalues as well.

\subsection{ C. Informational correlation in  systems with arbitrary initial state }
\label{Section:C}
The results obtained in Secs.\ref{Section:def} and \ref{Section:examples} 
are based on the tensor product initial state (\ref{inden}).
 If the initial state is more general, then eq.(\ref{TTmel2}) is not valid as 
 well as eq.(\ref{RITf}). In other words, the  $\varphi^A$-dependence may not 
 be collected in the density matrix $\rho^A(\varphi^A,0)$. In this case we also 
 may introduce  informational correlation $E^{AB}$ by eqs.(\ref{JA},\ref{EAB}). 
 {{} In turn, } 
the number $E^{AA}$ of  parameters  encoded into the subsystem $A$  
{{} may be introduced by eqs.(\ref{JB},\ref{EAA})}. Therewith
the vector $\hat X$ is defined by eq.(\ref{XT}) together with eqs.(\ref{XYZ}). 
Again, the number of parameters encoded into $\rho^A(\varphi^A,0)$ is defined 
by the multiplicity of the eigenvalues of $\rho^A(0)$.  However, the 
representation (\ref{RITex1})  for $X(\rho^B(\varphi^A,t))$,
$Y(\rho^B(\varphi^A,t))$ and $Z(\rho^B(\varphi^A,t))$ is not valid any 
more.  Inequality (\ref{Jfneq}) between  $E^{AB}$ and  $E^{AA}$ has no place as well. 
At first glance, this inequality must be replaced by the more formal one: 
\begin{eqnarray}\label{Jfneq2}
E^{AB}(t) \le E^{AA}.
\end{eqnarray}
However,  inequality (\ref{Jfneq2}) is not evident and might be wrong in general. In fact, applying the local transformation to the subsystem $A$ we influence on the whole density matrix $\rho(0)$
yielding the density matrix $\rho(\varphi^A,0)$ . However, only certain combinations of the elements of $\rho(\varphi^A,0)$ appear in   $\rho^A(\varphi^A,0)$.  Thus, some  of the parameters $\varphi_i$ might be missed from the local density matrix $\rho^A(\varphi^A,0)$, but might be detected in the whole density 
matrix $\rho(\varphi^A,0)$. This forces us to 
denote  the number of all parameters encoded into the initial density matrix  
$\rho(\varphi^A, 0)$ by $E^A$, $E^A\ge E^{AA}$.  {{} This quantity is defined }
by the equation (similar to eqs.(\ref{EAB}) and (\ref{EAA}))
\begin{eqnarray}
E^A(\varphi^A,0)= {\mbox{ran}} J(\rho(\varphi^A,0)) ,
\end{eqnarray}
where 
\begin{eqnarray}\label{JBAp}
J(\rho(\varphi^A,0))=\frac{\partial (\hat X_1(\rho(\varphi^A,0)),\dots, 
\hat X_{N^2-1}(\rho(\varphi^A,0))  }{\partial(\varphi_1,\dots,\varphi_{\tilde D^A})}.
\end{eqnarray}
Thus, there might be such parameters $\varphi_i$ that 
are not encoded into the initial reduced density matrix $\rho^A(\varphi^A,0)$, but might appear in the reduced density  matrix $\rho^B(\varphi^A,t)$ in the course of  evolution. 
The number of these parameters may not exceed the value $\delta E^{A}$, 
\begin{eqnarray}
\delta E^{A}(\varphi^A,t) = E^A(\varphi^A,t) - E^{AA}.
\end{eqnarray}
Consequently, instead of (\ref{Jfneq2}), the following inequality holds:
\begin{eqnarray}\label{Jfneq22}
E^{AB}(\varphi^A,t) \le E^{A}(\varphi^A,t).
\end{eqnarray}
Emphasize that $E^{AB}$ depends on $\varphi^A$ in the case of arbitrary initial state $\rho(0)$. It is obvious that  the normalized informational correlation  $E^{AB}_{norm}$ defined by formula (\ref{norm}) might be bigger then one in this case.

Now, let us calculate $E^{AB}$ using  eqs.(\ref{JA},\ref{EAB}).  In general,  the rank of the matrix $J(\rho^B)$ must be calculated numerically. 
For this purpose, we fix the time interval  $T_1<t<T_2$ taken for the parameter detection in the subsystem $B$ and introduce the set of  auxiliary functions  
\begin{eqnarray}\label{Mn}
&&
{\cal{M}}_n(t)= \frac{\hat M_n(t)}{\hat M_{n;max}},\;\;
\hat M_n(t) =
\int\limits_{ G^A} \sum_i |M_{ni}(\varphi^A,t)| d\Omega(\varphi^A),\;\;
\hat M_{n;max} =\max\limits_{T_1<t<T_2} \hat M_n(t),\\\nonumber
&&
n=1,\dots, \tilde D^A,
\end{eqnarray}
where $M_{ni}$ are the $n$th order minors of  $J(\rho^B)$, sum is over all minors, integration is over the whole $G^A$ and $\Omega(\varphi^A)$ is some measure. 
The function  ${\cal{ M}}_n(t)$ is positive if only at least some of the $n$th order minors $M_{ni}$ are nonzero on the non-zero volume subregion $g(t)$ of the region $ G^A$ (note that $g$ may depend on $t$). Then we define $E^{AB}(g(t),t)$ as the maximal order $n_0$ of the positive functions  ${\cal{ M}}_n(t)$, $n=1,\dots,n_0$, i.e., 
\begin{eqnarray}\label{Mn2}
E^{AB}(g(t),t) = \max\limits_{{{\cal{M}}_n(t)> 0}\atop{\varphi^A\in g(t)} }\, n(t) = n_0(t),
\end{eqnarray} 
so that $n_0$ depends on $t$ in general.
For the practical purpose, we might need to replace the positivity condition of ${\cal{M}}_n(t)$ by the following one:
\begin{eqnarray}\label{Mn3}
{\cal{M}}_{n}(t) \ge \varepsilon,
\end{eqnarray}
where $\varepsilon>0$ is some parameter predicted by  the errors of calculations and/or experiment.

Let us consider the case of stationary region $\hat g$, $\hat g\subset G^A$.
The time intervals suitable for the detection of the transfered parameters in the subsystem $B$ might be 
 defined numerically by the algorithm similar to that used in examples of Sec.\ref{Section:nonredEX2}.
First of all we 
  introduce  the auxiliary function ${{ M}}(t)$ defined as follows:
\begin{eqnarray}\label{hMap}
 M(t)=\frac{M^g_{n_0}(t)}{M^g_{max}},
\;\;
M^g_{n_0}(t)=\min_{\varphi^A\in  g} \sum_i |M_{n_0i}(\varphi^A,t)|,
\;\;
M^g_{max} = \max\limits_{T_1\le t \le T_2} M^g_{n_0}(t).
\end{eqnarray}
Formally, any time instant  corresponding to the positive $M$ is suitable for the parameter detection. However, 
if $M$ is too small, then there might be some obstacles for the correct detection of these parameters (for instance, fluctuations). Thus we take only such time subintervals 
 inside of the taken interval $T_1 < t < T_2$ that satisfy the following condition:
\begin{eqnarray}\label{Mpos}
M(t) > \tilde \varepsilon .
\end{eqnarray}
Here $ \tilde \varepsilon $ is some positive parameter, predicted by the required accuracy. For instance, $
\displaystyle \tilde \varepsilon =\frac{1}{2}$ in examples of Sec.\ref{Section:nonredEX2}.

In a similar way, we may  study the non-reducible correlations. The formulas (\ref{Mn}-\ref{Mpos}) hold with replacement $E^{AB} \to E^{AB;min}$, therewith  $M_{ni}$ must be the $n$th order minors of the matrix $H$ defined by eq.(\ref{HH}). 

%{{}
%Of particular interest is the case of LNUs, 
%when $E^{AA}=0$. These transformations for stationary systems have 
%been studied in \cite{F,GKB,DG}. But the problem of detection  of
%particular parameters $\varphi_i$ transfered from the subsystem $A$ to the subsystem $B$ has not been considered up to our knowledge.}

\end{document}